\documentclass[
aps
,prx
,amsmath,amssymb,amsfonts
,superscriptaddress
,floatfix
,longbibliography
,twocolumn,10pt
]{revtex4-1}

\usepackage{amsmath,amssymb,amsfonts}
\usepackage[usenames, dvipsnames]{color}
\usepackage[dvipsnames,svg justnames,x11names,hyperref]{xcolor}
\usepackage[toc,page]{appendix}
\usepackage[T1]{fontenc}
\usepackage[utf8]{inputenc}
\usepackage{times}
\usepackage{esint}
\usepackage[low-sup]{subdepth}

\usepackage{graphicx}
\usepackage{tabularx}
\newcolumntype{R}{>{\raggedleft\arraybackslash}X}
\usepackage{ragged2e}

\usepackage[
]{physpack}

\usepackage{pdfpages}
\makeatletter
\AtBeginDocument{\let\LS@rot\@undefined}
\makeatother

\makeatletter
\newcommand*{\balancecolsandclearpage}{%
  \close@column@grid
  \clearpage
  \twocolumngrid
}
\makeatother

\definecolor{linkcolor}{RGB}{6,69,173} % Wikipedia
\usepackage[
    unicode=true,
    pdfusetitle,
    bookmarks=false,
    colorlinks=true,
    linkcolor=linkcolor,
    urlcolor=linkcolor,
    citecolor=linkcolor,
    pdfencoding=auto]{hyperref}

\makeatletter
\@ifundefined{date}{}{\date{}}
\AtBeginDocument{
  
}
\makeatother

\begin{document}

\title{Real space thermalization of locally driven quantum magnets}

\author{Ronald Melendrez}
\affiliation{Department of Physics, Florida State University, Tallahassee, Florida 32306, USA}
\affiliation{National High Magnetic Field Laboratory, Tallahassee, Florida 32310, USA}

\author{Bhaskar Mukherjee}
\affiliation{Department of Physics and Astronomy, University College London, Gower Street, London WC1E 6BT, United Kingdom}

\author{Prakash Sharma}
\affiliation{Department of Physics, Florida State University, Tallahassee, Florida 32306, USA}
\affiliation{National High Magnetic Field Laboratory, Tallahassee, Florida 32310, USA}

\author{Arijeet Pal}
\affiliation{Department of Physics and Astronomy, University College London, Gower Street, London WC1E 6BT, United Kingdom}

\author{Hitesh J. Changlani}
\affiliation{Department of Physics, Florida State University, Tallahassee, Florida 32306, USA}
\affiliation{National High Magnetic Field Laboratory, Tallahassee, Florida 32310, USA}

\date{\today}

\begin{abstract}
The study of thermalization and its breakdown in isolated systems has led to a deeper understanding of non-equilibrium quantum states and their dependence on initial conditions. The role of initial conditions is prominently highlighted by the existence of quantum many-body scars, special athermal states with an underlying effective superspin structure, embedded in an otherwise chaotic many-body spectrum. Spin Heisenberg and $XXZ$ models and their variants in one and higher dimension have been shown to host exact quantum many-body scars, exhibiting perfect revivals of spin helix states that are realizable in synthetic and condensed matter systems. Motivated by these advances, we propose experimentally accessible, local, time-dependent protocols to explore the spatial thermalization profile and highlight how different parts of the system thermalize and affect the fate of the superspin.
We identify distinct parametric regimes for the ferromagnetic ($X$-polarized) initial state based on the interplay between the driven spin and the rest, including local athermal behavior where the driven spin effectively decouples, acting like a ``cold" spot while being instrumental in heating up the other spins. We also identify parameter regimes where the superspin remains resilient to local driving for long time scales. We develop a real and Floquet space picture that explains our numerical observations, and make predictions that can be tested in various experimental setups.
\end{abstract}

\maketitle
\section{Introduction}
In a set of pioneering papers ~\cite{Deutsch1991,Srednicki1994,Rigol2008,Jensen_Shankar} the question of thermalization of isolated quantum systems was posed sharply and addressed. It is now understood that generic isolated quantum systems satisfy the eigenstate thermalization hypothesis~\cite{Landau_Lifshitz}. Broadly said, local observables are insensitive to the choice of eigenstate at a given energy density and the system is ``self-thermal" i.e. it acts as its own heat bath. However, there are important exceptions, these include emergent integrable systems, for example, many-body localized systems~\cite{Basko2006, Pal2010, Oganesyan2007, Nandkishore2015, Abanin2019_Review}, and partially integrable systems or those with ``quantum scars".~\cite{Heller1984, Vafek2017, Shiraishi2017, Turner2018_np}
The search for quantum many-body scars (QMBS),
athermal states embedded in the spectra of otherwise
chaotic systems, has seen recent activity~\cite{Shiraishi2017,Turner2018_np,
Turner2018_prb, Moudgalya2018_Exact, Moudgalya2018_AKLT,
Lin2019, Khemani2019_RydbergIntegrable, Bull2019,
Schecter2019,Ok2019, Lee_PRBR2020, Lee_Pal_Changlani, McClartyscar, Wildeboer_scar, Pakrouski_scar, Voorden_scar, Spin1Kitaevscars, Chertkovscar, Hatsugai_scar, Song_2022, Sharma_spin1, Hirsch_scar, Serbyn_review, Chandran_review}
because of fundamental interest and due to proposals for using them for quantum
sensing~\cite{Dooley,Serbyn_review}. Though not expected for generic interacting systems~\cite{Ken_2022}, QMBS states do occur in realistic situations, especially when the Hilbert space is fragmented due to kinetic constraints \cite{sala2019ergodicity,khemani2019local,Lee_Pal_Changlani, Neupert_HSF, Moudgalya_HSF}. In a time-dependent setting, the presence of a global periodic drive can either destabilize or stabilize prethermal/athermal behavior associated with QMBS~\cite{Mukherjee_2020,zhao,Bluvstein_2021}, for example, under certain conditions the system can exhibit slow thermalization and dynamical freezing~\cite{freezing,AsmiPRXResonant}.

\begin{figure}
\includegraphics[width=\linewidth]{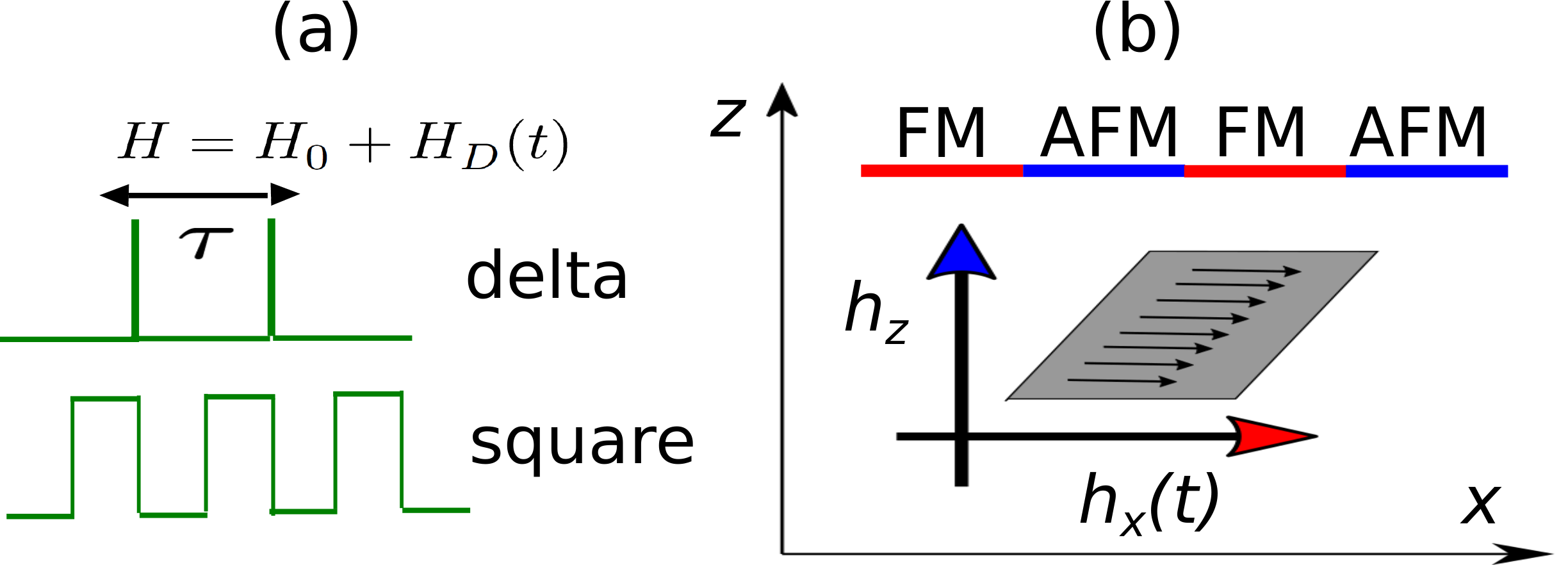}
\par  \vspace{0.5cm} 
\includegraphics[width=\linewidth]{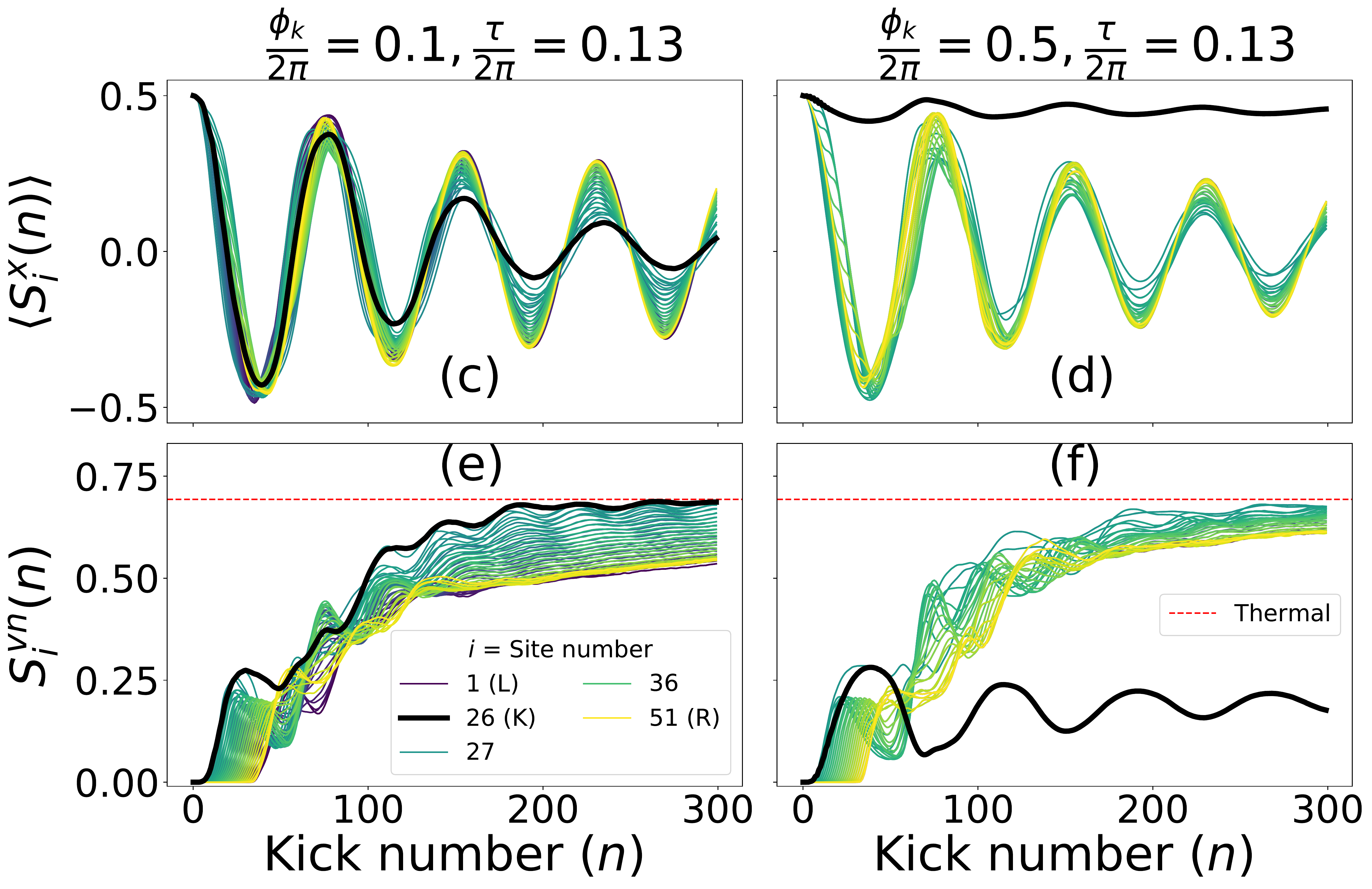}
\caption{Schematic of (a) local kick protocols and (b) the system studied in this work. The alternating (staggered) interactions place the ferromagnetic state in the middle of the spectrum. For the local delta kick, we show two characteristic behaviors for a $N=51$ site system with open boundary conditions with the kick at the central site (labelled by $K$). (c,d) show $\langle S^{x}_i\rangle$ and (e,f) the von Neumann entanglement entropy of representative sites. Results are for $J=1,h=0.1$,$\tau/2\pi = 0.13$ and (c,e) $\phi_k/2\pi = 0.1$ and (d,f) $\phi_k/2\pi =0.5$. The driven site thermalizes for weak kicks, for stronger kicks, it remains athermal on the time scale of observation. The numerical simulations were performed with TEBD, as discussed in the text. Labels $L$ and $R$ refer to the leftmost 
and rightmost edges of the chain respectively.}
\label{fig:schematic}
\end{figure}

In previous work, some of us identified the $XXZ$ model as a simple platform for realizing QMBS and Hilbert-space fragmentation (HSF)~\cite{Lee_PRBR2020, Lee_Pal_Changlani}. The model shares a common unifying theme with other models of scars, including the widely studied PXP one~\cite{Turner2018_np}; there is a ``superspin" whose precession is responsible for revivals in various numerically computed and experimentally measured physical observables.
Such a superspin can be realized as a ferromagnetic state 
embedded in the middle of the many-body energy spectrum (the exact SU(2) degeneracy being split by a magnetic field) by ``staggering" the
$XXZ$ model i.e. by alternating the sign of interactions on different geometric motifs~\cite{Lee_PRBR2020}. In one dimension this translates to a Hamiltonian with alternating nearest neighbor ferromagnetic and antiferromagnetic interactions. When the spins are prepared in a collective coherent state, for example in the $|X \rangle \equiv \Pi_i \bigotimes |\rightarrow \rangle_i$ state, and allowed to time evolve, their dynamics corresponds to that of a superspin. The interactions between the spins are rendered completely ineffective by the choice of initial conditions and there is no thermalization~\cite{Lee_PRBR2020}. However, this is a fine-tuned situation 
and one should generically expect thermalization when the system is perturbed.

\begin{figure*}
\includegraphics[width=\linewidth]{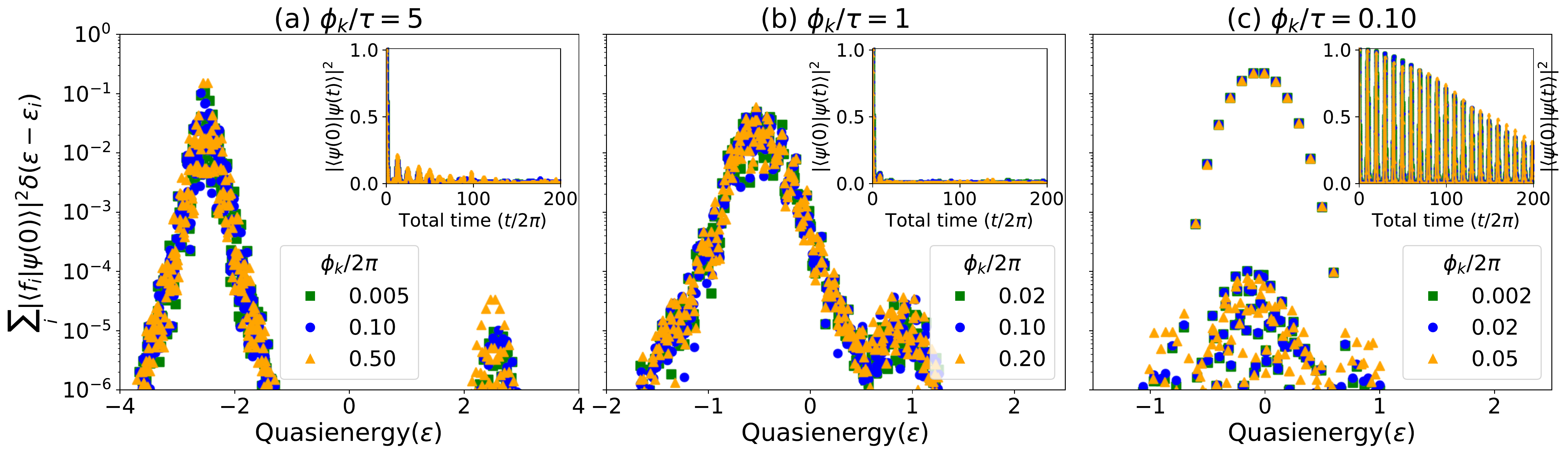}
\caption{Floquet overlap profile and (inset) survival probability/Loschmidt echo starting from the fully polarized $X$ state, for representative cases of the local delta kick protocol for $N=12$ sites with periodic boundary conditions with $J=1,h=0.1$ and (a) $\phi_k/\tau =5$ (b) $\phi_k/\tau=1$ (c) $\phi_k/\tau=0.1$. For ease of visualization (and to account for situations with near or exact degeneracies), quasienergies are binned with a spacing of $10^{-2}$, and the total overlap on the (nearly) degenerate manifold is reported.}
\label{fig:floquet_delta}
\end{figure*}

The question we ask here is: how do QMBS thermalize when subjected to time-dependent fields? Our motivation stems from the aim of expanding the existing dichotomy of classifying a system as either thermal or athermal. After all, could it be that there are parts of the system that are athermal (or prethermal on long time scales) while the rest have thermalized? We explore this question in the context of a periodic, local drive \cite{Thuberg,adhip,HubnerLocalDrive}, and demonstrate the crossover between regimes of weak HSF and quantum scarring where the system locally remains athermal (on the time scale of the observation) due to the
interplay between the dynamics of the driven spin and the rest of the system. We investigate two time-dependent potentials, schematically depicted in Fig.~\ref{fig:schematic}(a) and (b) whose similarities and differences we will highlight, especially in the context of the effective Floquet Hamiltonian they realize. In both cases the Hamiltonian for a $N$ site spin-1/2 chain is given by $H = H_{0}+ H_D(t)$ where, 
\begin{eqnarray}
	H_0 &\equiv& J\sum_{i=1}^{N} (-1)^i {\bf{S}_i} \cdot {\bf{S}_{i+1}} - h \sum_{i=1}^{N} S_i^z  
\end{eqnarray}
and $H_D(t)$ is a time-dependent drive term. ${\bf{S}_i} \equiv (S^{x}_i,S^{y}_i,S^{z}_i)$ refer to the usual spin-1/2 operators on site $i$, 
$J$ (set to 1 throughout) is the alternating (staggered) ferro- and antiferromagnetic interaction strength and $h$ is the strength of the magnetic
(Zeeman) field. $i+1$ is taken modulo
$N$ for periodic boundary conditions. For open boundary
conditions the index $i$ on the first sum runs from 1 to $N-1$.
Unlike the (Bethe ansatz) integrable uniform Heisenberg chain~\cite{Bethe_1931}, $H_0$ is known to be non-integrable~\cite{Mukherjee_Melendrez_2024}. Its study, primarily in the context of its ground state properties, has a long history due to its relevance to Haldane spin chains~\cite{HaldaneSpin1,Hida1992AlternatingChain,KohmotoAlternating}. 

For the first ``delta kick" protocol, the drive term is
\begin{eqnarray}
H_D(t) &\equiv& - \sum_{n>0} \sum_i \phi_i S_i^x \delta(t-n\tau)
\label{eq:drive_delta}
\end{eqnarray}
where $\phi_i$ denotes the strength of the 
applied transverse (direction taken to be $x$) 
magnetic field strength on site $i$.  
The delta kick can be thought of as a reasonable approximation  
to the situation where the duration of the transverse field pulse is much shorter than $\tau$ and other time scales associated with $H_0$ (e.g. $1/J$ and $1/h$).
While there is a considerable body of work on models where a single large spin is kicked~\cite{BennakerKicked,haake1987classical,Sinha_2021KickedDicke}, 
we emphasize that $H_0$ here has only local interactions and the system behaves as a collective spin (superspin)
degree of freedom only for certain initial conditions.
We consider the case where only a single spin at site $k$ is driven with strength $\phi_k$, while the rest of the spins are undriven.

The second drive protocol is that of a symmetric square pulse,
\begin{eqnarray}
H_D(t) &\equiv& \sum_i \gamma_i \textrm{Sgn} \Big(\sin \Big(\frac{2\pi t}{\tau} \Big) \Big)S^x_i
\label{eq:drive_square}
\end{eqnarray}
where $\gamma_i$ is the strength of the transverse field on site $i$. 
Like the delta kick protocol, we will consider here only the case of a single driven spin.
We note that a similar pulse protocol, but with different $H_0$, 
was studied in the context of global drives (all spins driven)  
to demonstrate the existence of resonant scars~\cite{AsmiPRXResonant}. 

For both drive protocols, we have used the time-dependent 
$\langle S_i^{x} \rangle$ and onsite von Neumann entanglement entropy to identify 
regimes where the driven spin either collectively thermalizes with the rest of the spins, as in Fig.~\ref{fig:schematic} (c,e), 
or essentially disentangles itself from the remainder of the spins as in Fig.~\ref{fig:schematic}(d,f). 
The latter case serves as an example of a system that is locally kept athermal (or "cold") by driving whereas the rest of the system "heats up" and thermalizes. 
In the rest of the paper we will explore these phenomena further.

\section{Floquet and real space picture of thermalization} 
In this section we discuss both the Floquet ("quasienergy") space and local (real space) picture of the phenomena shown 
in Fig.~\ref{fig:schematic} by placing their behavior in the context of the familiar picture of QMBS. We make the notion of the Floquet quasienergy spectrum precise and clarify how it assumes the role of eigenenergies of the time-independent Hamiltonian, where identification of an isolated manifold with large overlap on the initial state revealed the existence of QMBS~\cite{Turner2018_np, Turner2018_prb}. 

\subsection{Floquet Hamiltonian and Floquet overlap profile}
For stroboscopic times, defined to be any time which is an integer multiple of the drive periodicity, the unitary Floquet operator completely encodes all information about the time evolution. For a single period, the Floquet operator is given by 
\begin{equation}
F(\tau) \equiv e^{-iH_F \tau} = \sum_j e^{-i \epsilon_j \tau} |f_j \rangle \langle f_j|
\end{equation} 
where $H_F$ is defined to be the effective Hermitian Floquet Hamiltonian, $|f_j\rangle$ is its $j^{th}$ eigenvector and $\epsilon_j$ is the corresponding quasienergy. For the two drive protocols introduced in Eqns.~\eqref{eq:drive_delta} and ~\eqref{eq:drive_square} we have,
\begin{eqnarray} 
F(\tau) &=& e^{+i \sum_i \phi_i S_i^x } e^{-i H_0 \tau} \;\;\;\;\;\;\;\;\;\;\;\;\;\;\;\;\;\;\;\;\;\;\;\;\;\;\;\;\;\textrm{delta} \label{eq:floquet_delta} \\
F(\tau) &=& e^{-i(H_0-\sum_i\gamma_i S_i^x)\frac{\tau}{2}} e^{-i(H_0+\sum_i\gamma_i S_i^x)\frac{\tau}{2}} \;\;\;\textrm{square} \label{eq:floquet_square} 
\end{eqnarray}
Knowing $|f_j \rangle$ and $\epsilon_j$ and hence $c_j \equiv \langle f_j | \psi(0) \rangle$, enables inference of many properties of the dynamics, for example, the survival probability (Loschmidt echo) is 
\begin{equation}
\Big|\langle \psi(0) | \psi(n\tau) \rangle \Big|^2 = \sum_{j,k} |c_j|^2 |c_k|^2 e^{-i n(\epsilon_j-\epsilon_k)\tau}.
\end{equation}
where $n \geq 0$ is an integer. Since the quasienergies typically have spacings that statistically resemble those that arise from a random matrix, 
this quantity generally goes to zero in the long time limit by virtue of the superposition of the (almost) random phases. However, this is not always the case - a regularity of the quasienergy spacings for states with dominant $|c_i|$ leads to robust revivals of the survival probability and other observables. 

\begin{figure*}
\includegraphics[width=0.55\linewidth]{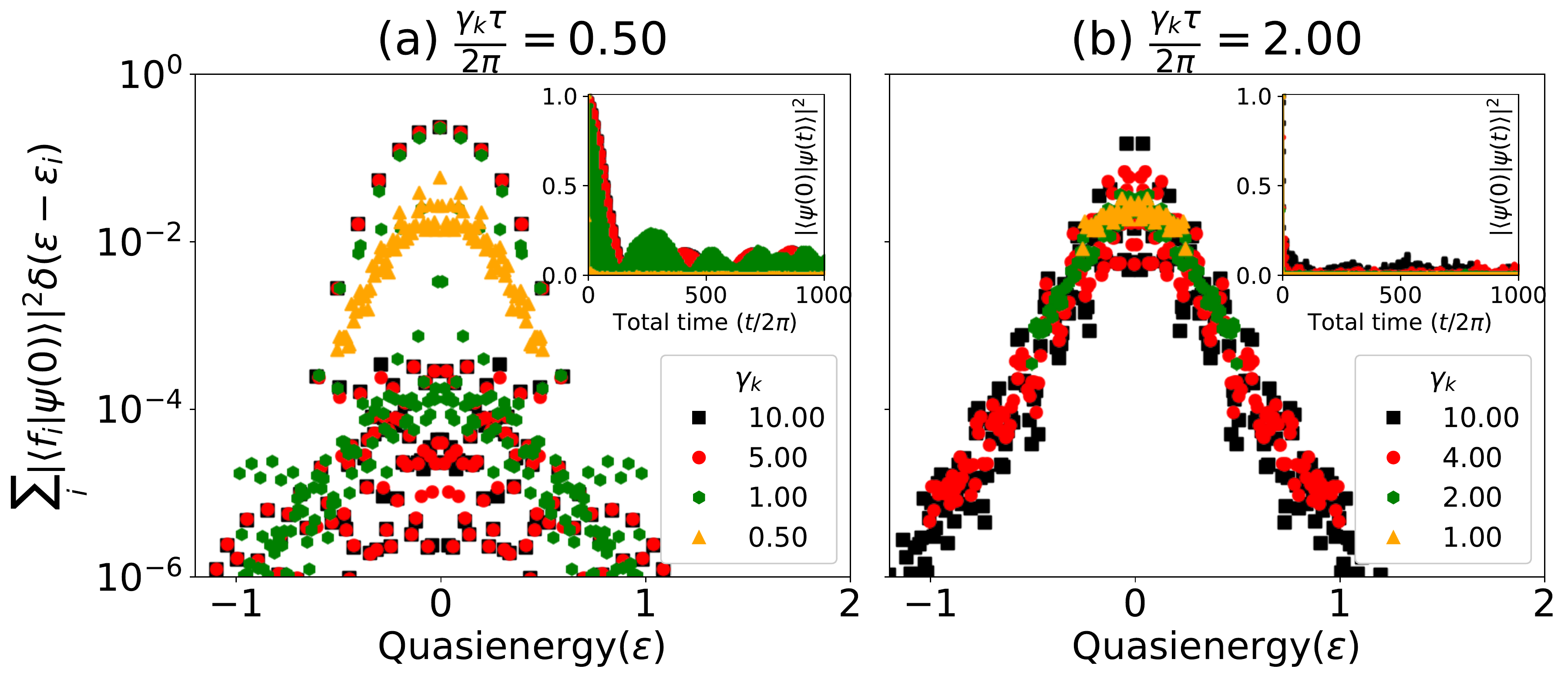}
%\par  \vspace{0.5cm} 
\includegraphics[width=0.44\linewidth]{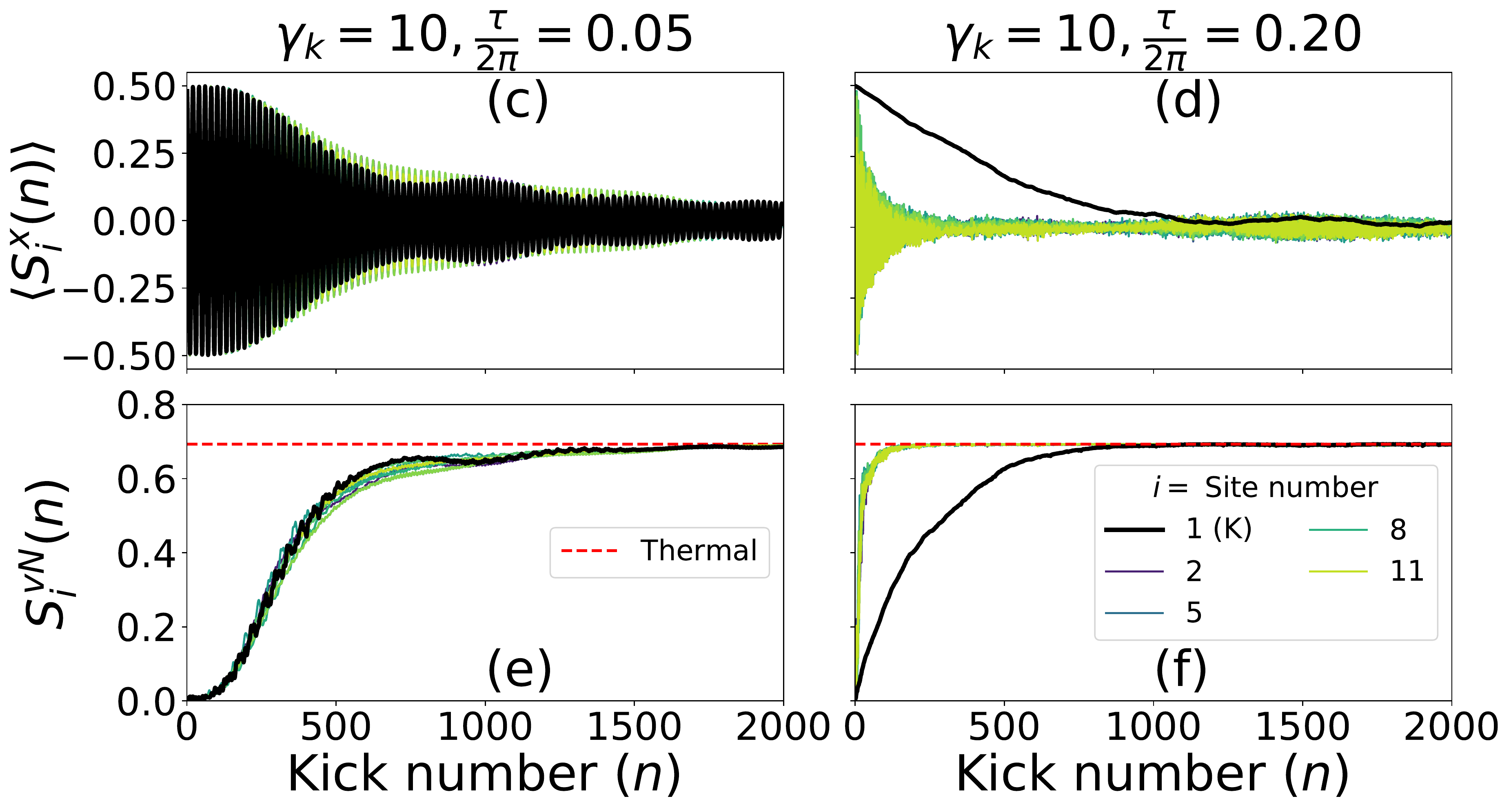}
\caption{Panels (a,b) show the Floquet overlap profile and (inset) survival probability/Loschmidt echo starting from the $|X\rangle$ state, for representative parameters of the local square pulse protocol for $N=12$ sites with periodic boundary conditions and $J=1,h=0.1$. $\gamma_k \tau / 2\pi$ was fixed to the value of (a) $0.50$ and (b) $2$, while individually varying $\gamma_k$ and $\tau$. The value in (b) satisfies the resonant condition, see text. Quasienergies are binned with a spacing of $10^{-2}$, and the total overlap on the (nearly) degenerate manifold is reported. Panels (c,d) show $\langle S_i^x (t) \rangle$ and panels (e,f) show the von Neumann entanglement entropy $S_i^{vn}(t)$ for all sites ($i$) for the cases of $J=1, h=1$ and (c,e) $\gamma_k=10, \tau / 2\pi = 0.05$ and (d,f) $\gamma_k=10, \tau/2\pi = 0.20$. 
Panels (c,e) show that a single site drive is sufficient to thermalize all the sites. Panels (d,f) show slow relaxation of local properties at the driven site due to the resonant condition. Label $K$ is used to indicate the driven site.
}
\label{fig:floquet_square}
\end{figure*}

The quasienergies (eigenenergies of $H_F$) of the driven system 
play the role analogous to energies of the corresponding undriven
system. The overlap of the initial state on to the eigenvectors of the Floquet
operator illuminates which states participate in the time evolution of the system.
We thus use the plot of $|c_j|^2$ as a function of $\epsilon_j$ as a diagnostic tool and refer to it as the ``Floquet overlap profile", as in  Figs.~\ref{fig:floquet_delta} and \ref{fig:floquet_square}. To smoothen the appearance of certain features in the plots, especially in situations with exactly (or nearly) degenerate quasienergies, 
we divide the quasienergy space into small bins and report the total overlap of all states in the bin.  

\subsection{Delta kick}
Consider first the case of the local delta kick. On performing a Baker-Campbell-Hausdorff (BCH) expansion of Eq.~\eqref{eq:floquet_delta}, the Floquet Hamiltonian, to lowest order in the kick strength is, 
\begin{equation}
	H^{(0)}_F = H_0 - \frac{\phi_k}{\tau} S^{x}_k,
\label{eq:HF_delta}
\end{equation}
as shown in Appendix~\ref{app:1}. Physically, one can think of the second term as a local magnetic field 
of strength $\phi_k/\tau$ in the $x$ direction, a viewpoint that will be useful for interpreting the numerical results in the next section. 

The mathematical form of Eq.~\eqref{eq:HF_delta} suggests that the Floquet overlap profile and survival probability must be approximately a 
function of $\phi_k/\tau$. In Fig.~\ref{fig:floquet_delta} we explore the validity of this assertion by plotting the Floquet overlap profile (main panels) and the Loschmidt echo (inset) for three representative values of $\phi_k/\tau$ showing multiple $\phi_k$ for each. We find that, for the range of $\phi_k/\tau$ plotted here, our data collapse nicely on top of each other, especially for lower values of $\phi_k$. Expectedly, this data collapse begins to break down for large $\phi_k$ due to the presence of higher order terms in the Floquet Hamiltonian. (Also note that $\phi_k = 2\pi\times \textrm{integer}$ is special and corresponds to the situation where there is no kick). 

We now explain the trends seen in Fig.~\ref{fig:floquet_delta}. If there were no $H_0$, the Floquet quasienergies are $-\phi_k/2\tau$ 
and $+\phi_k/2\tau$, which correspond to Floquet eigenvectors that have $S^{x}_{k} = +1/2$ and $S^{x}_{k} = -1/2$ respectively. (
The $|X\rangle$ state has overlap only onto the manifold of states with $S^{x}_{k}=+1/2$.) 
When $H_0$ is present, but $\phi_k/\tau$ is large, $S^{x}_{k}$ is not an exact integral of motion, just an approximate one. This situation 
can be viewed as the realization of weak HSF, two weakly coupled fragments (with $S^{x}_k \approx \pm 1/2$) appear in Hilbert space. 
The $|X\rangle$ state now has overlap with both fragments, the larger overlaps come from states with $S^{x}_{k} \approx + 1/2$. Importantly, the Floquet 
overlap profile now has ``bands" centered at $\pm \phi_k/2 \tau$ which have spread out. When these bands do not overlap, as in Fig.~\ref{fig:floquet_delta}(a) 
which corresponds to the case of $\phi_k/\tau=5$, the kicked spin is athermal. %(This value of $\phi_k/\tau$ also corresponds approximately to the parameters shown in Fig.~\ref{fig:schematic} where the entanglement entropy of the kicked site is much lower than the rest). 
This local physics is not apparent in the Loschmidt echo, plotted in the inset of Fig.~\ref{fig:floquet_delta}(a); this quantity does not 
show any prominent oscillations, and instead decays rapidly. This is not unexpected; the Loschmidt echo is a global property of the state 
which decays with time because majority of the spins lose coherence. This thermalization of the undriven spins arises due to the absence 
of any regular structure in the Floquet overlap profile; there is no manifold of states with large overlaps and regular spacing of quasienergies.

On making $\phi_k/\tau$ smaller, the previously separated Floquet bands broaden further and eventually begin to merge with one another, 
as is seen in Fig.~\ref{fig:floquet_delta}(b). When this happens, the driven spin does not act significantly differently from the rest of the spins. 
This collective thermalization is also corroborated by the results in Fig.~\ref{fig:schematic}. When the local delta kick is weakened even further, 
a long thermalization scale emerges due to the vicinity to a "perfect scar" since the $|X\rangle$ state is in the null space of the staggered Heisenberg term - it has a decomposition onto the tower of $2(N/2)+1 = N+1$ states. The Floquet overlap profile for $\phi_k/\tau=0.1$ in Fig.~\ref{fig:floquet_delta}(c) demonstrates this very clearly, the quasienergy spacing between states with non-zero overlap is $h$ and it arises from the Zeeman splitting of the embedded ferromagnet. %(The limit $\phi_k \rightarrow 0 $ corresponds to the undriven case i.e. $H_F = H_0$.)

\subsection{Symmetric square pulse}
The Floquet framework, coupled with either the BCH expansion of the
participating operators or the Floquet-Magnus (F-M) expansion, offers a way 
to understand, and hence engineer, the lifetime of QMBS states. For example, 
consider a driven spin with a kick that alternates in sign during one time 
period. In the limit of extremely rapid driving (high frequency driving), 
the kicks of opposite sign ``effectively cancel out" and the system is 
essentially undriven. Hence, the frequency and strength of the drive 
can be used as knobs for controlling decoherence times - higher frequency and 
weaker drive strengths favor longer decoherence times. 

The symmetric square pulse that we consider here also alternates in sign in
one time period. As shown in Appendix~\ref{app:2}, the Floquet Hamiltonian 
for the local square pulse to lowest order is, 
\begin{widetext}
\begin{eqnarray}
    H_F^{(0)}&=&(-1)^k[S^x_kS^x_{k+1}+\frac{2\sin(\frac{\gamma_k \tau}{2})}{\gamma_k \tau}(S^y_kS^y_{k+1}+S^z_kS^z_{k+1})-\frac{2(1-\cos(\frac{\gamma_k \tau}{2}))}{\gamma_k \tau}(S^z_kS^y_{k+1}-S^y_kS^z_{k+1})]\nonumber\\
    &&-(-1)^k[S^x_{k-1}S^x_k+\frac{2\sin(\frac{\gamma_k \tau}{2})}{\gamma_k \tau}(S^y_{k-1}S^y_k+S^z_{k-1}S^z_k)-\frac{2(1-\cos(\frac{\gamma_k \tau}{2}))}{\gamma_k \tau}(S^z_{k-1}S^y_k-S^y_{k-1}S^z_k)]\nonumber\\
	&&-\frac{2h}{\gamma_k \tau}(\sin(\frac{\gamma_k  \tau}{2})S^z_k+(1-\cos(\frac{\gamma_k \tau}{2}))S^y_k)+\sum_{\substack{i=1\\i\neq (k-1,k)}}^{N}(-1)^{i} {\bf{S}_i} \cdot {\bf{S}_{i+1}}-
    h\sum_{\substack{i=1\\i\neq k}}^{N}S^z_i
\label{eq:HF_square}
\end{eqnarray}
\end{widetext}
Two important takeaways from Eq.~\eqref{eq:HF_square} are (1) $H_F^{(0)}$ just depends on $\gamma_k \tau$, a finding confirmed by the (approximate) data collapse seen in the Floquet overlap profile in Fig.~\ref{fig:floquet_square} (a,b) and (2) there are special values of drive frequency and strength - the so called ``resonant condition"~\cite{AsmiPRXResonant} ($\gamma_k \tau = 4\pi \times \textrm{integer}$)  - where the system is locally athermal, a phenomenon referred to as dynamic freezing. Since the F-M expansion is most accurate at high drive frequency ($\gamma_k \tau \ll 1$), we expect the Floquet Hamiltonian to be well approximated by $H_F^{(0)}$ only in that regime. 

At high frequencies ($\gamma_k \tau \ll 1$), the drive is effectively rendered ineffective and $H^{(0)}_F \approx H_0$. 
In this regime, the $|X\rangle$ state is an almost perfect QMBS - it shows coherent oscillations in the survival probability and 
other observables (not shown). At intermediate frequencies $\gamma_k \tau \sim 1$, the Floquet overlap profile, as in Fig.~\ref{fig:floquet_square}(a), 
continues to show the characteristic manifold associated with QMBS and the associated superspin. However, there are are now subdominant modes 
that are important for thermalization at long times. (The data collapse for fixed $\gamma_k \tau$ is most accurate for large $\gamma_k$ and small $\tau$.) 
Fig.~\ref{fig:floquet_square} (c,e) shows the time dependent value of the onsite $\langle S_i^{x} \rangle$ and the von Neumann entanglement entropy; there are collective coherent oscillations in both quantities at short times, but eventual relaxation at long times. Thus, single site driving is sufficient to thermalize all the sites including the driven site, which is reminiscent of the behavior of the delta-kicked system for intermediate $\phi_k/\tau$. 

Figures~\ref{fig:floquet_square}(b,d,f) correspond to the case of a 
resonant frequency. The Floquet overlap profile shows that the 
characteristic QMBS manifold is lost, instead the situation resembles that 
of the delta-kick for large $\phi_k/\tau$. Unlike the case of a global drive at resonance~\cite{AsmiPRXResonant}, 
the locally driven spin is not entirely frozen out but it does relax significantly slowly compared to the other undriven spins. 
This ruining of dynamic freezing is caused by higher order terms in the F-M expansion of $H_F$, which some of us have pursued in detail 
elsewhere~\cite{Mukherjee_Melendrez_2024}. 

\section{Real-space, real-time profile of thermalization}
In this section we take a more refined look at the real-space, real-time picture of thermalization 
for a chain with open boundaries, where the central site is subject to a periodic delta kick. 
We simulate the short time dynamics of a $N=51$ site system with the matrix product state (MPS) 
based time evolving block decimation (TEBD) algorithm~\cite{Vidal_TEBD} employing a maximum bond dimension of $400$. 
To address (extremely) long time behavior, we simulate a 
much smaller system size of $N=13$ with exact diagonalization.

More generally we find that for a given set of drive parameters, the Floquet dynamics can be sensitive to $N$, an issue that 
we address in Appendix~\ref{app:3}. (That said, we have observed that the $N=31$ open chain exhibits similar qualitative behavior as $N=51$ for the same drive parameters.).

\subsection{Short time dynamics with TEBD}
\begin{figure}
\includegraphics[width=\linewidth]{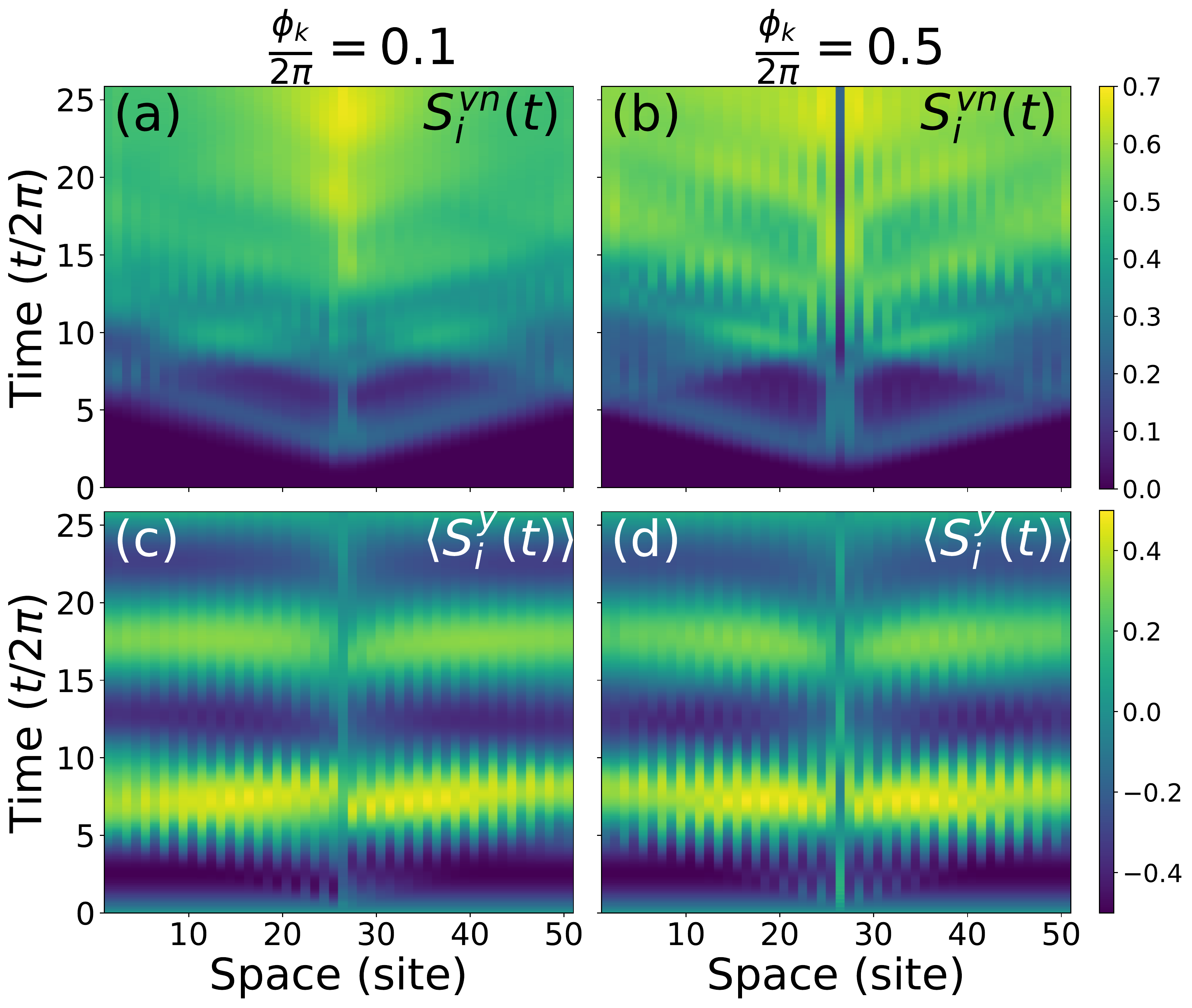}
\caption{Space-time thermalization profile for a $N=51$ site delta kicked staggered Heisenberg chain with open boundary conditions, showing the von Neumann entanglement entropy (a,b) $S_i^{vn} (t)$ of each site and (c,d) $\langle S_i^{y}(t)\rangle$. Space and time correspond to the horizontal and vertical axes respectively and the color represents the value of the physical quantity. Results are for $J=1,h=0.1,\tau/2\pi = 0.13$ and (a,c) $\phi_k/2\pi = 0.1$ and (b,d) $\phi_k/2\pi =0.5$.}
\label{fig:realspace1}
\end{figure}

To visualize dynamics of $N=51$ spins in both space and time we construct a ``space-time" plot, as in Fig.~\ref{fig:realspace1}, 
space is shown horizontally and time is shown vertically and the color represents 
the value of the quantity being investigated. Since the system is initially prepared in the $|X\rangle$ state, 
the von Neumann entanglement entropy of each spin is exactly zero to begin with. 
As time progresses, the entanglement spreads out in a "cone" and distant regions begin 
to feel the effects of the periodic kicking. This can be seen prominently in Fig.~\ref{fig:realspace1} for $\tau/2\pi = 0.13$ 
for both $\phi_k/2\pi = 0.1$(panel a) and $\phi_k/2\pi=0.5$ (panel b) at short times. (A time step of $\tau/20$ was used for the TEBD simulations, 
smaller time steps gave similar results). After this initial phase, the regions around the central spin show a 
prominent dip in their entanglement - they get cold after an initial phase of heating up - which 
appears as two blue lobes around the kicked spin. These regions eventually heat 
up at longer times, more generally, the plots show oscillatory behavior both in 
space and time.~\footnote{We also note that similar observations, albeit for a different model and
observables, have been reported recently in Ref.~\cite{Duan_scar}.} 
(Note that the system is not inversion symmetric because of the 
alternating $J$'s, but the plot of the entanglement entropy appears to be 
approximately symmetric about the central site).  
As pointed out earlier in Fig.~\ref{fig:schematic}, the driven site acts as a cold spot for $\phi_k/2\pi = 0.5$ - visually it appears 
as a blue vertical line in the space-time plot in Fig.~\ref{fig:realspace1}(b). 

To further probe how the driven spin behaves relative to the other (undriven) ones, we plot $\langle S^{y}_i \rangle$ for 
the two parameter sets in Figures~\ref{fig:realspace1}(c) and (d). On the time scale plotted, $\langle S^{y}_i\rangle$ shows collective revivals 
but they decay with time. Many of its features, at short times, can be understood without worrying about the interactions ($J$ terms). 
For example, for small $\phi_k$, following Eq.~\eqref{eq:HF_delta} the additional local effective magnetic field contribution along the $x$ direction is small. 
Thus the driven spin, like the other spins, precesses predominantly in the $x-y$ plane i.e. the axis about which it precesses is close to the 
direction of the applied field ($z$ direction). Since the superspin picture is intact, the driven spin is (largely) in phase with the others; 
this can be seen for the case of $\phi_k/2\pi = 0.1$ in Fig.~\ref{fig:realspace1}(c). This picture changes when the drive strength is large - 
the axis about which the driven spin precesses is now much closer to the $x$-axis, however, the axis for the other spins is unaffected. 
Since the starting state is the $|X\rangle$ state, with only a small component of the spin orthogonal to the axis of precession, 
the strength of the oscillations in $\langle S^{y}_k (t) \rangle$ are weak. These oscillations are also out of phase with respect to those of 
the other spins, as can be seen in Fig.~\ref{fig:realspace1}(d). 

\begin{figure}
\includegraphics[width=\linewidth]{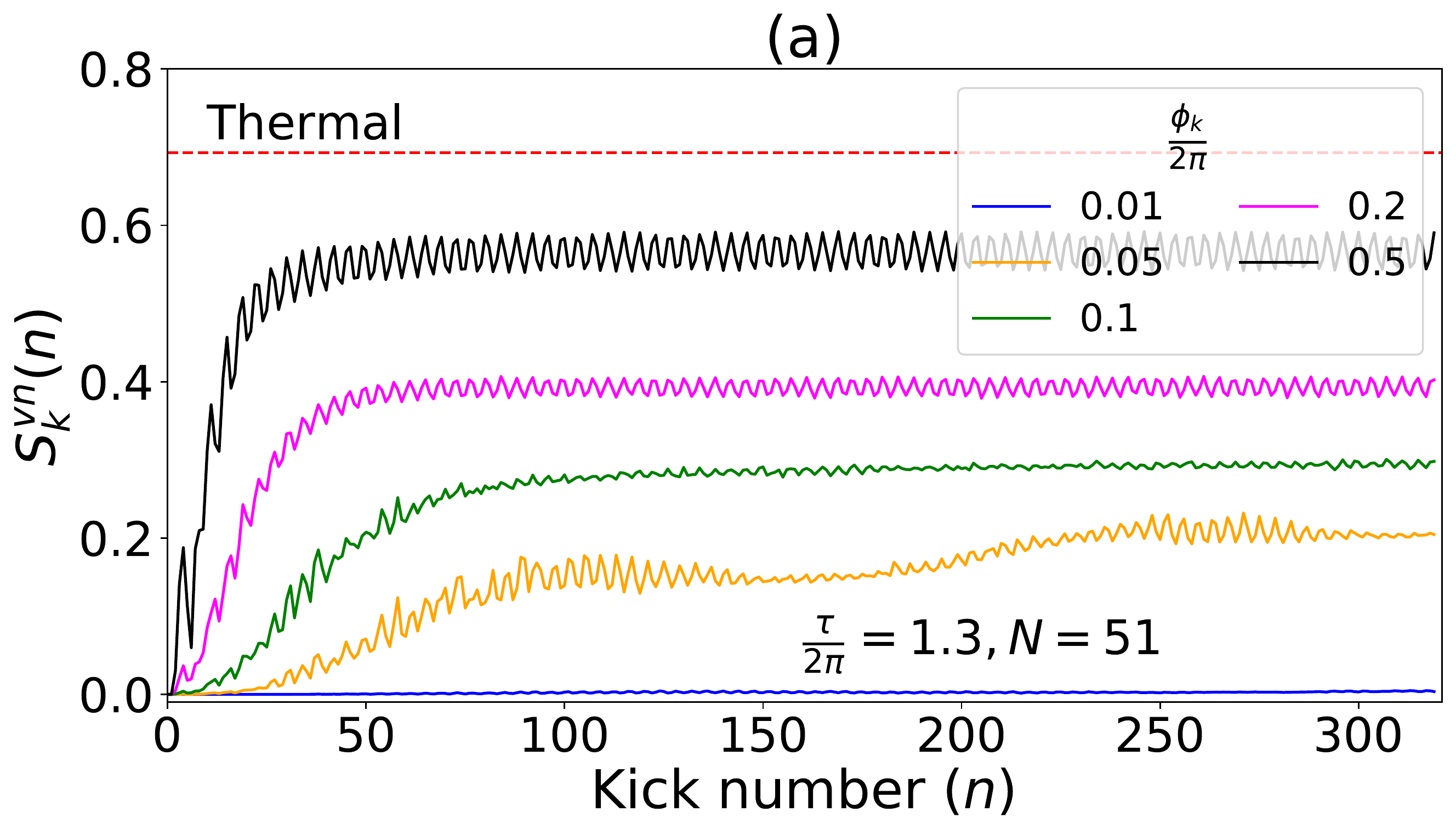}
\par  \vspace{0.3cm} 
\includegraphics[width=\linewidth]{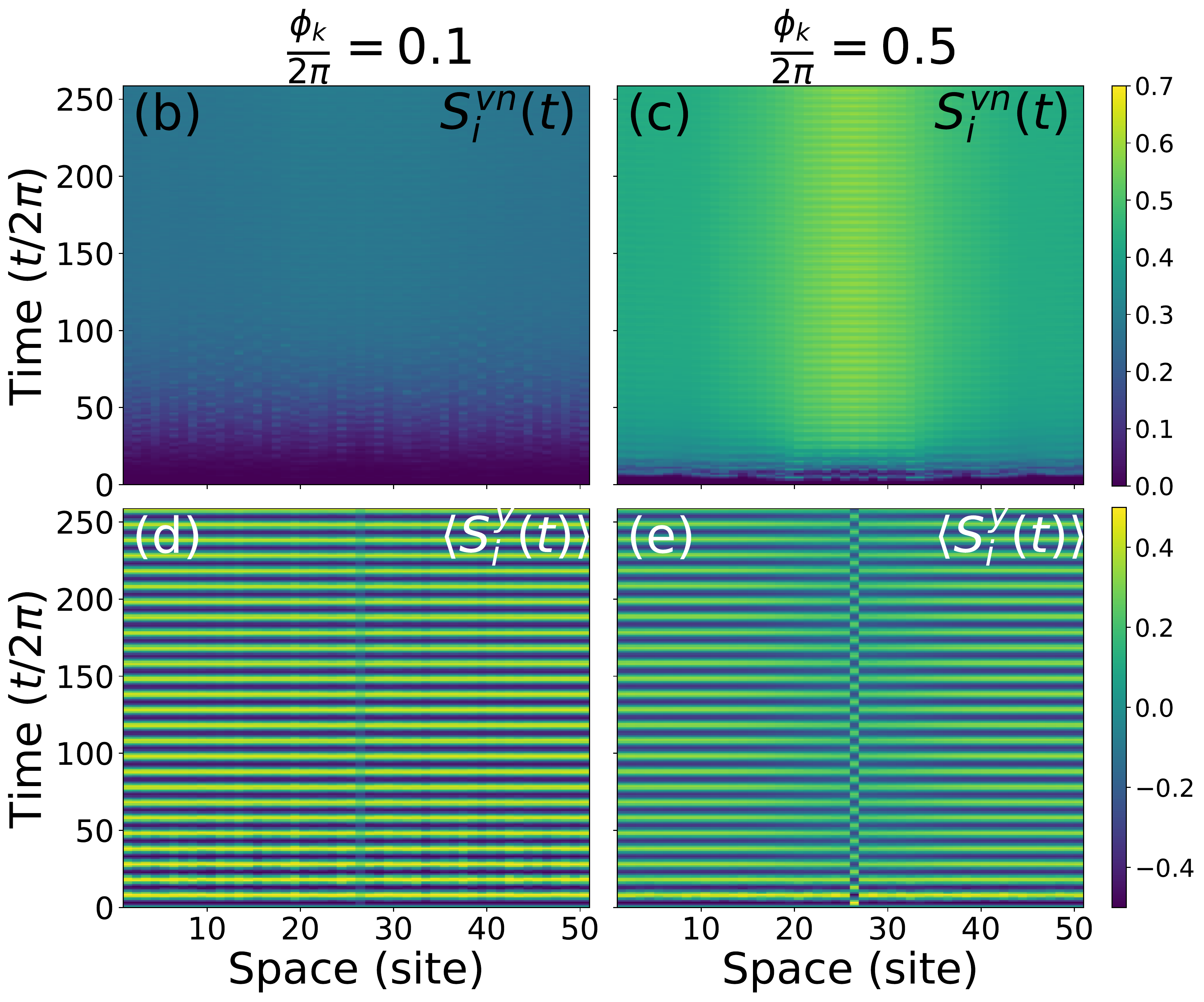}
\caption{ Panel (a) shows TEBD results for the time-dependent von Neumann entanglement entropy of the central kicked site for the $N=51$ open chain and $\tau/2\pi = 1.30$ and a few representative values of $\phi_k/2\pi$. The red dashed line corresponds to the maximal entropy of $\ln 2$. For the same system and drive frequency as in (a), the lower panels show space-time plots of the Von Neumann entanglement entropy (panels b,c) and the expectation value of $\langle S^{y}_i (t)\rangle$ (panels d,e) for two values of drive strength $\phi_k/2\pi = 0.1$ (panels b,d) and $\phi_k/2\pi=0.5$ (panels c,e).}
\label{fig:realspace2}
\end{figure}

For the $N=51$ site chain, we also study the case of
effectively weaker kicks by fixing the kick frequency to a value 10 times 
smaller i.e. $\frac{\tau}{2\pi}=1.30$. A time step of $\tau/200$ was used for the TEBD calculations, 
i.e. the same overall time step as that for the previously discussed cases. 
Fig.~\ref{fig:realspace2} (a) shows the time dependence of the von Neumann 
entanglement entropy for the driven spin for a few representative values of 
$\phi_k$ for this value of $\tau$. After an initial stage of entanglement 
growth, the entanglement entropy of the kicked spin (and other spins as seen
in the space-time plot in panels (b) and (c)) fluctuates about a non zero 
value; the value of this short-time ``plateau" and 
the strength of the oscillations both grow with $\phi_k$. The fact that this 
value is far from the thermal expectation of $\ln 2$, and that it is 
similar for all spins, %suggests that despite the breaking of translational
%symmetry, the superspin picture is intact even after the information from the kicked site has reached the boundaries of the system. 
reflects the resilience of the superspin to local driving, at least at short 
times.

Panels (d) and (e) of Fig.~\ref{fig:realspace2} show $\langle S^{y}_i (t) \rangle$ for the two values of $\phi_k$. We observe robust revivals in both cases on the time scale of the simulation, and the in-phase versus out of phase behavior of the driven spin closely parallels that seen for the case of lower $\tau$ in Fig.~\ref{fig:realspace1}. However, for $\phi_k = \pi$, the driven spin is not decoupled or ``cold", on the contrary it is slightly more entangled compared to rest of the spins.

\subsection{Long time dynamics with exact diagonalization}
Unfortunately, the TEBD simulations reveal information only about the short-time dynamics. Thus it is unclear whether our observations correspond to truly athermal or just transient prethermal behavior. To address this issue, we have explored the possibility of similar trends for smaller systems where full diagonalization of the Floquet operator is possible which, in turn, allows arbitrary stroboscopic times to be accessed. 

\begin{figure}
\includegraphics[width=\linewidth]
{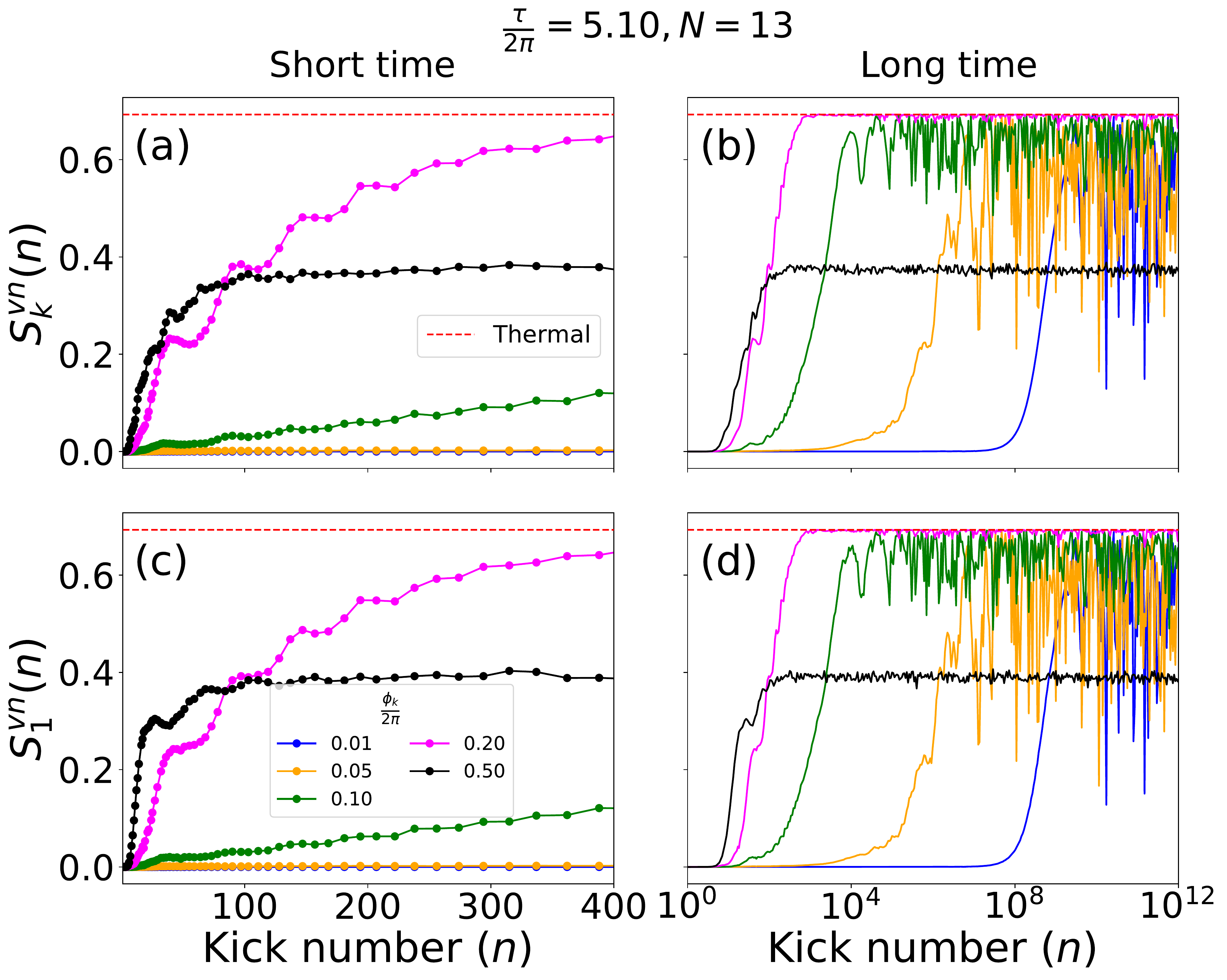}
\caption{Exact diagonalization results for the time-dependent von Neumann entanglement entropy of the central kicked site (panels a,b) and the leftmost boundary spin (site 1, panels c,d) for the $N=13$ open chain for $\tau/2\pi = 5.10$ and a few representative values of $\phi_k/2\pi$. The parameters used were $J=1,h=0.1$. 
The red dashed line corresponds to the maximal entropy of $\ln 2$. Panels (a,c) correspond to short time scales and are plotted on a linear axis and panels (b,d) correspond to long time scales plotted on a log axis. Markers correspond to stroboscopic times at which the quantity was evaluated, the connecting lines are guides to the eye.}
\label{fig:realspace3}
\end{figure}

Figures ~\ref{fig:realspace3}(a) and (b) show our results for the von Neumann entanglement entropy of the kicked site on a logarithmic time grid at short and long times (spanning many orders of magnitude) respectively for the $N=13$ open chain and $\tau/2\pi=5.10$. Note that the value of $\tau$ was chosen to be different in comparison to the $N=51$ case. (The sensitivity of the Floquet dynamics on system size is discussed in Appendix~\ref{app:3}) %This is because $H_0 \tau$ enters the Floquet operator - the number of terms in $H_0$ scale linearly with system size - and thus a meaningful comparison between different sizes for the same $\phi_k$ is possible only if $\tau$ is rescaled. 
Panels (c) and (d) show that the leftmost boundary spin closely follows the central/kicked spin at both short and long times like the other spins in the system (not shown).

At short times, the dynamics of the small chain qualitatively resembles that of the larger chain. There are some differences however, for example, the entropy curves for $\phi_k/2\pi=0.2$ and $\phi_k/2\pi=0.5$ cross. At much longer times, well beyond those accessible to TEBD, the entropy crosses over to significantly higher values, but not always the maximal value of $\ln 2 \approx 0.693$.  This is most prominently seen for the case of $\phi_k = \pi$ where the value is relatively constant (at approximately 0.4) over multiple decades of time. As expected, the time scale at which the rapid crossover from low to high entanglement occurs is inversely proportionate to $\phi_k$. Additionally, for small values of $\phi_k$, we observe visibly large fluctuations even at very long times indicating some remnant coherence, a feature that we intend to explore more systematically elsewhere. The deviation from the maximal entropy suggests the possible existence of truly athermal behavior or alternatively, a very long prethermal time scale~\cite{Abanin2017}, for some of the parameter sets studied here.

We emphasize that the existence of finite entanglement plateaus, for example those seen in Figs.~\ref{fig:realspace2} and ~\ref{fig:realspace3}, is considerably different from the case where the system is not driven. In the latter case, the system always remains in a (zero entanglement) product state at all times during the evolution. As noted earlier, when there is no drive, the interactions are rendered ineffective due to the choice of initial condition~\cite{Lee_Pal_Changlani}. However, in the weakly driven case, interactions facilitate the initial period of entanglement growth and the system reaches a quasi-steady state where it collectively remains as a superspin for a time scale that is $\phi_k$ dependent. In the extremely long time limit, this superspin typically, but not always, loses its coherence completely leading to thermalization.

The possible explanation of this effect is rooted in a consideration of the Floquet Hamiltonian. In the limit that $\phi_k/\tau$ is small, the zeroth 
order term may be inadequate for even a qualitatively accurate description~\cite{Mukherjee_Melendrez_2024}. Instead, one must consider higher order 
terms which may either compete or collaborate to renormalize the effective magnetic field that the kicked site experiences, as seen in 
Eqns.~\eqref{eq:HF_0_delta} and ~\eqref{eq:HF_1_delta}. If the effective field is driven to a small value, thermalization is slow. 
Consequently, the system may also exhibit a thermalization rate that sensitively depends on the size of the system $N$. 
On the one hand, the periodic kick has the effect of disturbing the coherently precessing spins, so larger $N$ leads to longer times for the
entire system to feel the disturbance. On the other hand, larger $N$ also means there are more channels to exchange energy which could lead to faster 
thermalization. These competing effects manifest themselves in non-trivial ways and we have addressed this issue numerically in Appendix~\ref{app:3}. 
In particular, we have provided evidence of non-monotonic (with $N$) effects with the help of Fig.~\ref{fig:specific_tau_phi_various_N} and Fig.~\ref{fig:tau_scan_N_9_11_13}. 

\section{Conclusion}
In summary, we have explored the different dynamical regimes of a locally driven staggered (alternating sign) 
Heisenberg spin chain for local drive protocols that should be experimentally accessible. For the $X$-polarized initial state, the driven spin crosses 
over from full thermalization to athermal dynamics, effectively decoupling from the rest of the spins, 
as the rest of the system thermalizes. Both drive protocols showed this effect, and the numerical 
observations were explained within the framework of the lowest order Floquet Hamiltonian. We also 
discovered parametric regimes where a quasi steady state is reached after a period of initial entanglement growth, 
with collective oscillations, suggesting the resilience of the superspin picture for effectively weak drives.
An important conceptual outcome is that both the local and Floquet space pictures are complementary for understanding the mechanism 
by which the system thermalizes or fails to do so. More generally, our analyses 
also potentially carry over to the case of periodic drives with spatial deformed Hamiltonians~\cite{Wen_2022}. 

Given the simplicity of the model and drive protocols investigated here, we believe these predictions can be tested. 
There are now many synthetic realizations of spin and Bose Hubbard systems~\cite{Jad}: in addition to Rydberg atoms, ytterbium-171 has been recently 
used to realize an effective transverse Ising model~\cite{Tan_TFIM_2021} and hyperfine states of lithium realize $XXZ$ models with tunable anisotropy~\cite{Jepsen2022}. Our observation of the local breakdown of thermalization in an otherwise thermal system for a certain initial state provides a non-trivial mechanism for protecting information in a periodically driven system and sheds light on the novel behavior of dynamics of many-body entanglement. 

\section*{Acknowledgements} 
We thank Kyungmin Lee for useful discussions in the initial stages of the project and for a previous collaboration. R.M., P.S. and H.J.C. acknowledge support from Grant No. NSF DMR- 2046570 and Florida State University (FSU) and the National High Magnetic Field Laboratory. The National High Magnetic Field Laboratory is supported by the National Science Foundation through Grant No.s~DMR-1644779, DMR-2128556 and by the state of Florida. A.P. and B. M. were funded by the European Research Council (ERC) under the European Union's Horizon 2020 research and innovation programme (Grant Agreement No. 853368). We also thank the Planck cluster and the Research Computing Center (RCC) at FSU for computing resources.
The MPS based TEBD calculations were based on the ITensor library.~\cite{ITensor}

\appendix 

\section{$H_F$ for the periodic delta kick protocol}
\label{app:1}
In this Appendix we derive an approximation for $H_F$ for the periodic delta kick protocol with the help of the truncated 
BCH expansion applied to Eq.~\eqref{eq:floquet_delta}. %In what follows we present expressions only 
%for chains with periodic boundary conditions, the expressions for chains with open boundaries 
%differ in the limits of the sum over the bond terms.

Set $X \equiv i\phi_k S^x_k$, $Y \equiv -iH_0\tau$ and $Z \equiv -iH_F\tau$, 
then $H_F$ is given by
\begin{eqnarray}
H_F&=&\frac{iZ}{\tau} = \frac{i}{\tau}\ln(e^Xe^Y)
\end{eqnarray}
Defining $H_F \equiv \sum_{n} H_F^{(n)}$, we determine $H_F^{(n)}$ using the BCH formula and get,
\begin{widetext}
\begin{subequations}
\begin{eqnarray}
H_F^{(0)}&=&\frac{i}{\tau}(X+Y)=H_0-\frac{\phi_k }{\tau}S^x_k \label{eq:HF_0_delta}\\
H_F^{(1)}&=&\frac{i}{\tau}\frac{[X,Y]}{2}=\frac{\phi_k }{2} \Big( (-1)^k \Big( S^y_{k-1}S^z_k+
S^y_kS^z_{k+1}-S^z_{k-1}S^y_k-S^z_kS^y_{k+1} \Big) -hS^y_k \Big) \label{eq:HF_1_delta}
%H_F^{(2)}&=&\frac{i}{\tau}\frac{1}{12}([X,[X,Y]]-[Y,[X,Y]])\nonumber\\
%&=&
%\frac{\phi ^2}{12}(hS^z_1+S^y_1S^y_2+S^z_1S^z_2)-\frac{\phi \tau}{24}(-%S^x_2+S^x_1(1+2h^2+4hS^z_2)-2S^y_1(S^y_2S^x_3-S^x_2S^y_3)-2S^z_1(S^z_2S^x_3-S^x_2(S^z_3-2h)))\nonumber \\
%H_F^{(3)}&=&\frac{i}{\tau}\frac{-[Y,[X,[X,Y]]}{24}\nonumber\\
%&=&\frac{\phi ^2\tau}{24}(S^z_1(S^y_2S^x_3-S^x_2S^y_3)+S^y_1(-S^z_2S^x_3+S^x_2(S^z_3-2h)))\nonumber\\
%H_F^{(4)}&=&\frac{i}{\tau}(\frac{[Y,[Y,[Y,[X,Y]]]]}{720} - \frac{[X,[X,[X,[X,Y]]]]}{720} - \frac{[X,[Y,[Y,%[X,Y]]]]}{360} + \frac{[Y,[X,[X,[X,Y]]]]}{360} + \frac{[Y,[X,[Y,[X,Y]]]]}{120} -\nonumber\\
%&&\frac{[X,[Y,[X,[X,Y]]]]}{120})
\end{eqnarray}
\end{subequations}
\end{widetext}

\section{$H_F$ for the square pulse protocol}
\label{app:2}
In this Appendix we provide a detailed calculation of the stroboscopic $H_F$ for the square pulse protocol using the Floquet-Magnus (F-M)
expansion in a rotating frame. The F-M method yields $H_F$ as a 
perturbative expansion in inverse drive frequency ($\tau$). However, transition to a 
rotating frame automatically ensures an infinite order resummation in $\tau$. Thus, we obtain a series expansion of $H_F$ in inverse drive-amplitude ($1/\gamma_k$) where all the terms are resummed in $\tau$ \cite{asen, replica, anatoli1, anatoli2}. This extends the validity of $H_F$ to the low drive frequency regime but limits it to the high drive amplitude regime.

%The Hamiltonian for the driven system is given by, 
%\begin{equation}
%    H(t)=H_0+H_D(t)
%\end{equation}
As mentioned in the text, the local drive at the $k$-th site is given by $H_D(t)=\gamma_k \textrm{Sgn}(\sin(\omega t))S^x_k$ with $\omega=2\pi/\tau$. It is convenient to transform the time-dependent Hamiltonian $H(t) = H_0 + H_D(t)$ to a rotating frame as follows,
\begin{equation}
    H_{rot}(t)=W^{\dagger}H(t)W-i W^{\dagger}\partial_tW
\end{equation}
where 
\begin{equation}
    W(t)=e^{-i\int_0^tdt'H_D(t')}=e^{-i\theta(t)S^x_k}
\end{equation}
and
\begin{equation} 
\theta(t)=\gamma_k t\Theta(\tau/2-t)+\gamma_k (\tau-t)\Theta(t-\tau/2).
\end{equation}
This transformation removes the driving term ($H_D(t)$) from the 
Hamiltonian in rotating frame ($H_{rot}(t)$) and we get
\begin{widetext}
\begin{eqnarray}
    H_{rot}(t)&=&W^{\dagger}H_0W \nonumber \\ 
            &=& \mathcal{H}(k-1,k)+\mathcal{H}(k,k+1)-h(\cos(\theta)S^z_k+\sin(\theta)S^y_k)+
	\sum_{\substack{i=1\\i\neq (k-1,k)}}^{N}(-1)^{i}{\bf{S}_i} \cdot {\bf{S}_{i+1}}-
    h\sum_{\substack{i=1\\i\neq k}}^{N}S^z_i 
\end{eqnarray}
where 
\begin{equation}
\mathcal{H}(k,k+1)=(-1)^k[S^x_kS^x_{k+1}+\cos(\theta)(S^y_kS^y_{k+1}+S^z_kS^z_{k+1})-\sin(\theta)(S^z_kS^y_{k+1}-S^y_kS^z_{k+1})]
\end{equation}
\end{widetext}
and we have used
\begin{subequations}
\begin{eqnarray}
W^{\dagger}S^z_kW &=&\cos(\theta)S^z_k+\sin(\theta)S^y_k \\
W^{\dagger}S^y_kW &=& -\sin(\theta)S^z_k+\cos(\theta)S^y_k
\end{eqnarray}.
\end{subequations}

%The stroboscopic Floquet Hamiltonian in F-M expansion 
%is given by 
%\begin{equation}
%    H_F=\sum_{n=0}^{\infty}H_F^{(n)}
%\end{equation}
%We calculate upto 2nd order ($n=2$).
%\subsubsection{0 th order F-M}
The zeroth order Floquet Hamiltonian is just the time averaged $H_{rot}(t)$ over one time period.
\begin{widetext}
\begin{eqnarray}
    H_F^{(0)}&=&\frac{1}{\tau}\int_0^\tau H_{rot}(t)dt
  =\mathcal{H}_F^{(0)}(k-1,k)+\mathcal{H}_F^{(0)}(k,k+1)+h_F^{(0)}(k)+
	\sum_{\substack{i=1\\i\neq (k-1,k)}}^{N}(-1)^{i} {\bf{S}_i} \cdot {\bf{S}_{i+1}}-
    h\sum_{\substack{i=1\\i\neq k}}^{N}S^z_i
\end{eqnarray}
where 
\begin{subequations}
\begin{eqnarray}
\mathcal{H}_F^{(0)}(k,k+1)&=&(-1)^k[S^x_kS^x_{k+1}+\frac{2\sin(\frac{\gamma_k \tau}{2})}{\gamma_k \tau}(S^y_kS^y_{k+1}+S^z_kS^z_{k+1})-\frac{2(1-\cos(\frac{\gamma_k \tau}{2}))}{\gamma_k \tau}(S^z_kS^y_{k+1}-S^y_kS^z_{k+1})] \\
h_F^{(0)}(k)&=&-\frac{2h}{\gamma_k \tau}[\sin(\frac{\gamma_k \tau}{2})S^z_k+(1-\cos(\frac{\gamma_k \tau}{2}))S^y_k]
\end{eqnarray}
\end{subequations}
\end{widetext}

\section{Non-monotonic system size dependence of Floquet dynamics} 
\label{app:3}
In this Appendix we provide numerical evidence for the sensitivity of Floquet dynamics to the system size $N$. 
We consider chains with odd $N$ and open boundaries, with the central site being driven/kicked periodically. The spins are initialized, at $t=0$, to the $|X \rangle$ state. 

\begin{figure}
\includegraphics[width=\linewidth]{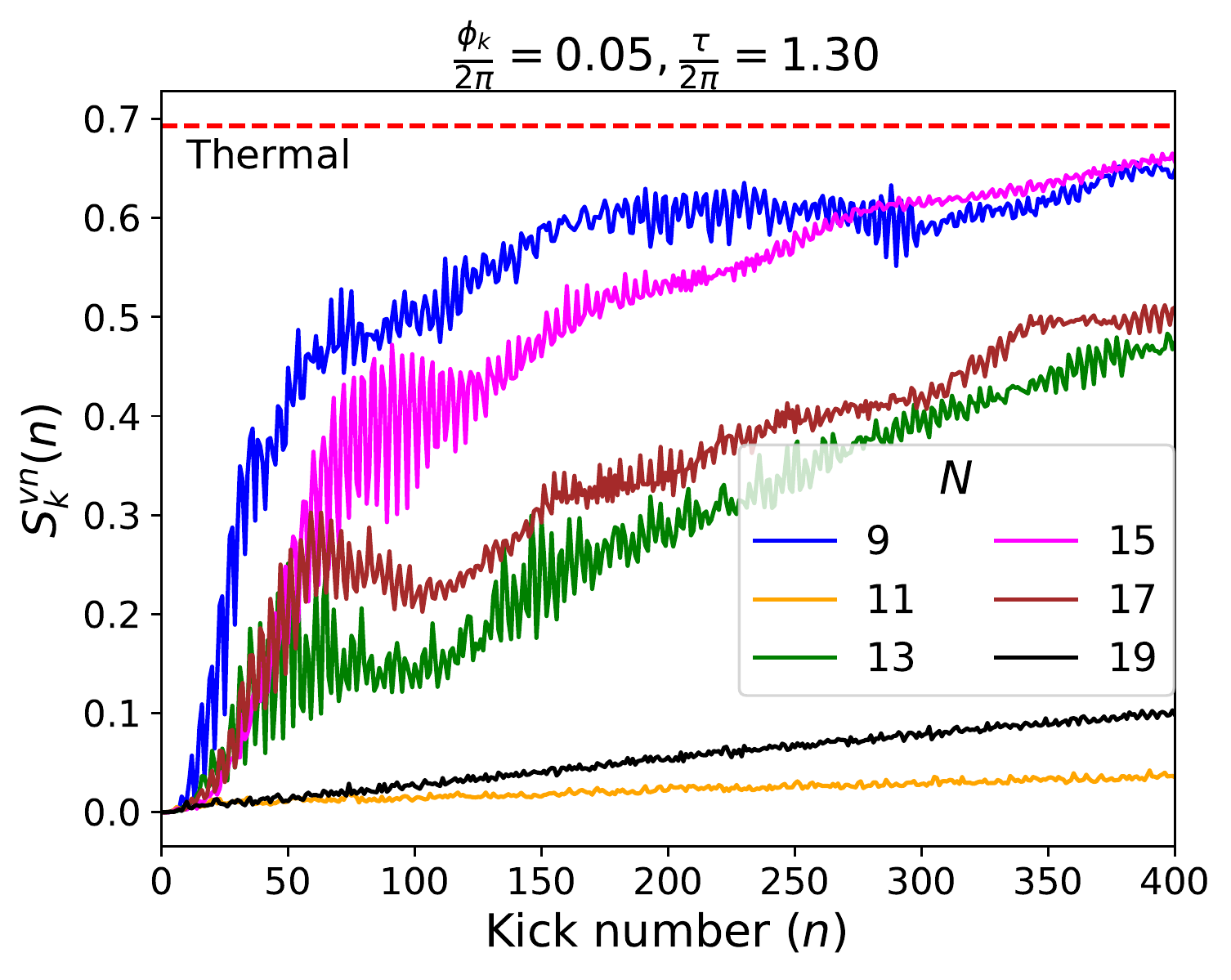}
\caption{Time dependent von Neumann entanglement entropy of the driven (kicked) site, denoted by $S^{vn}_k$, for chains of length $N$ with open boundary conditions, whose central site is subject to a periodic delta kick. The parameters are $J=1,h=0.1, \frac{\phi_k}{2\pi} = 0.05$ and $\frac{\tau}{2\pi} = 1.30$.}
\label{fig:specific_tau_phi_various_N}
\end{figure}

In Fig.~\ref{fig:specific_tau_phi_various_N} we plot the von Neumann entanglement entropy for the driven (central) site at stroboscopic times for $\phi_k/2\pi=0.05$ and $\tau/2\pi=1.30$ and $J=1,h=0.1$. The results were obtained with TEBD with a maximum bond dimension of 512, and thus there is no truncation error for $N\le 19$. The time step was chosen to be small ($\delta t=\tau/200$), such that the TEBD results matched the exact diagonalization results (for $N=9,11,13$) to a high accuracy, throughout the entire time window shown in the plot.

We observe non-monotonic (with $N$) trends. The $N=11$ chain shows almost no change in the entanglement entropy from its initial value of zero, in sharp contrast to the $N=9$ case. The time profile then changes again for larger $N$, the rate of increase of entanglement entropy increases for $N=13$ and $N=15$ but then slows down again for $N=17$ and (more prominently) for $N=19$. 

\begin{figure}[h]
\includegraphics[width=\linewidth]{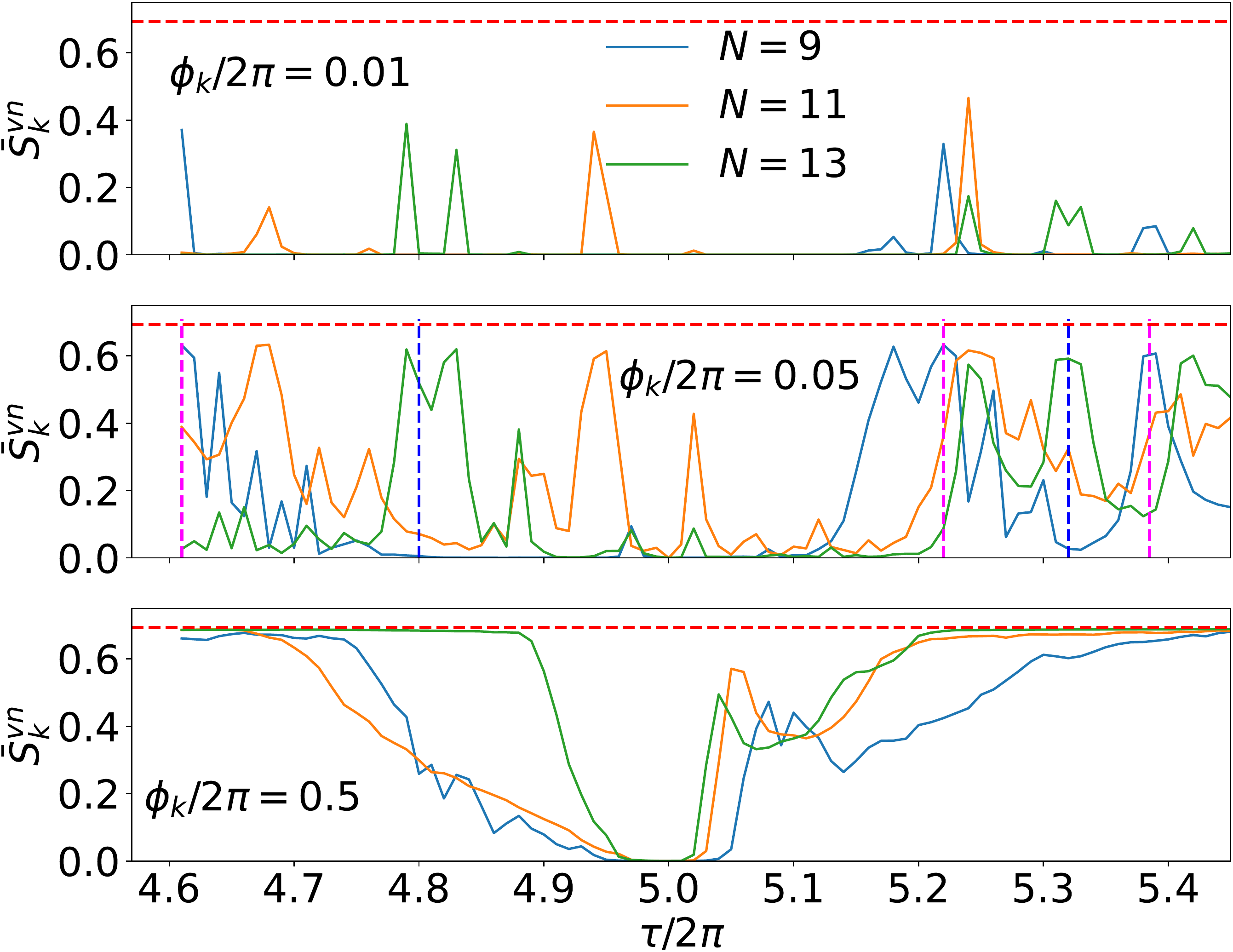}
\caption{Entanglement entropy of the kicked site, averaged over first 1000 drive cycles (${\bar S^{\text{vn}}_k}$) vs $\tau/2\pi$ for three different system sizes. Horizontal red dashed line in each panel denotes the thermal value of entropy ($\ln 2$). Vertical dashed lines in the middle panel are guide to the eye for the cases when the relaxation becomes faster (blue) and slower (magenta) with system size. The parameters are $J=1,h=0.1$.}
\label{fig:tau_scan_N_9_11_13}
\end{figure}

To further explore this effect, in Fig.~\ref{fig:tau_scan_N_9_11_13} we plot the average entanglement entropy of the kicked spin over the first 1000 drive cycles (kicks), for $N=9,11,13$, computed with exact diagonalization. We consider a range of $\tau/2\pi$, in the vicinity of $5.10$, in steps of $0.01$, and representative values of $\phi_k/2\pi = 0.01, 0.05$ and $0.5$ with $J=1, h=0.1$. This plotted quantity captures the average (initial) rate of relaxation towards thermal equilibrium - larger values indicate fast approach towards the thermal state and low values indicate slow or no relaxation in the time window of the first 1000 kicks. 
Note that the case of $\tau/2\pi=5$ is special given that $h=0.1$ - it corresponds to half the time period of the precession. This means that the delta kick is given to a spin when it is in an eigenstate of $S^{x}_k$, which has the effect of only contributing a global phase to the time-dependent wavefunction, which does not disturb the coherent precession of the spins. Thus the state remains a product state for this $\tau$.

For $\phi_k/2\pi=0.01$ the relaxation is, in general, slow for most $\tau$. However, there are some $\tau$ for which the relaxation rate is faster, importantly they occur at different $\tau$ for different $N$. This is confirmed in the middle panel, for $\phi_k/2\pi = 0.05$, the vertical dashed lines indicate situations where the relaxation becomes prominently faster or slower with increasing system size. The bottom panel shows our results for $\phi_k/2\pi = 0.5$. Once again there is generically a strong dependence on $N$, however, there are regimes (for example, $\tau/2\pi \approx 5.1$) where the relaxation rates for the three sizes shown here are coincidentally similar.

\bibliography{refs}

%merlin.mbs apsrev4-1.bst 2010-07-25 4.21a (PWD, AO, DPC) hacked
%Control: key (0)
%Control: author (0) dotless jnrlst
%Control: editor formatted (1) identically to author
%Control: production of article title (0) allowed
%Control: page (1) range
%Control: year (0) verbatim
%Control: production of eprint (0) enabled
\begin{thebibliography}{71}%
\makeatletter
\providecommand \@ifxundefined [1]{%
 \@ifx{#1\undefined}
}%
\providecommand \@ifnum [1]{%
 \ifnum #1\expandafter \@firstoftwo
 \else \expandafter \@secondoftwo
 \fi
}%
\providecommand \@ifx [1]{%
 \ifx #1\expandafter \@firstoftwo
 \else \expandafter \@secondoftwo
 \fi
}%
\providecommand \natexlab [1]{#1}%
\providecommand \enquote  [1]{``#1''}%
\providecommand \bibnamefont  [1]{#1}%
\providecommand \bibfnamefont [1]{#1}%
\providecommand \citenamefont [1]{#1}%
\providecommand \href@noop [0]{\@secondoftwo}%
\providecommand \href [0]{\begingroup \@sanitize@url \@href}%
\providecommand \@href[1]{\@@startlink{#1}\@@href}%
\providecommand \@@href[1]{\endgroup#1\@@endlink}%
\providecommand \@sanitize@url [0]{\catcode `\\12\catcode `\$12\catcode
  `\&12\catcode `\#12\catcode `\^12\catcode `\_12\catcode `\%12\relax}%
\providecommand \@@startlink[1]{}%
\providecommand \@@endlink[0]{}%
\providecommand \url  [0]{\begingroup\@sanitize@url \@url }%
\providecommand \@url [1]{\endgroup\@href {#1}{\urlprefix }}%
\providecommand \urlprefix  [0]{URL }%
\providecommand \Eprint [0]{\href }%
\providecommand \doibase [0]{http://dx.doi.org/}%
\providecommand \selectlanguage [0]{\@gobble}%
\providecommand \bibinfo  [0]{\@secondoftwo}%
\providecommand \bibfield  [0]{\@secondoftwo}%
\providecommand \translation [1]{[#1]}%
\providecommand \BibitemOpen [0]{}%
\providecommand \bibitemStop [0]{}%
\providecommand \bibitemNoStop [0]{.\EOS\space}%
\providecommand \EOS [0]{\spacefactor3000\relax}%
\providecommand \BibitemShut  [1]{\csname bibitem#1\endcsname}%
\let\auto@bib@innerbib\@empty
%</preamble>
\bibitem [{\citenamefont {Deutsch}(1991)}]{Deutsch1991}%
  \BibitemOpen
  \bibfield  {author} {\bibinfo {author} {\bibfnamefont {J.~M.}\ \bibnamefont
  {Deutsch}},\ }\bibfield  {title} {\enquote {\bibinfo {title} {Quantum
  statistical mechanics in a closed system},}\ }\href {\doibase
  10.1103/PhysRevA.43.2046} {\bibfield  {journal} {\bibinfo  {journal} {Phys.
  Rev. A}\ }\textbf {\bibinfo {volume} {43}},\ \bibinfo {pages} {2046}
  (\bibinfo {year} {1991})}\BibitemShut {NoStop}%
\bibitem [{\citenamefont {Srednicki}(1994)}]{Srednicki1994}%
  \BibitemOpen
  \bibfield  {author} {\bibinfo {author} {\bibfnamefont {Mark}\ \bibnamefont
  {Srednicki}},\ }\bibfield  {title} {\enquote {\bibinfo {title} {Chaos and
  quantum thermalization},}\ }\href {\doibase 10.1103/PhysRevE.50.888}
  {\bibfield  {journal} {\bibinfo  {journal} {Phys. Rev. E}\ }\textbf {\bibinfo
  {volume} {50}},\ \bibinfo {pages} {888} (\bibinfo {year} {1994})}\BibitemShut
  {NoStop}%
\bibitem [{\citenamefont {Rigol}\ \emph {et~al.}(2008)\citenamefont {Rigol},
  \citenamefont {Dunjko},\ and\ \citenamefont {Olshanii}}]{Rigol2008}%
  \BibitemOpen
  \bibfield  {author} {\bibinfo {author} {\bibfnamefont {Marcos}\ \bibnamefont
  {Rigol}}, \bibinfo {author} {\bibfnamefont {Vanja}\ \bibnamefont {Dunjko}}, \
  and\ \bibinfo {author} {\bibfnamefont {Maxim}\ \bibnamefont {Olshanii}},\
  }\bibfield  {title} {\enquote {\bibinfo {title} {Thermalization and its
  mechanism for generic isolated quantum systems},}\ }\href {\doibase
  10.1038/nature06838} {\bibfield  {journal} {\bibinfo  {journal} {Nature
  (London)}\ }\textbf {\bibinfo {volume} {452}},\ \bibinfo {pages} {854}
  (\bibinfo {year} {2008})}\BibitemShut {NoStop}%
\bibitem [{\citenamefont {Jensen}\ and\ \citenamefont
  {Shankar}(1985)}]{Jensen_Shankar}%
  \BibitemOpen
  \bibfield  {author} {\bibinfo {author} {\bibfnamefont {R.~V.}\ \bibnamefont
  {Jensen}}\ and\ \bibinfo {author} {\bibfnamefont {R.}~\bibnamefont
  {Shankar}},\ }\bibfield  {title} {\enquote {\bibinfo {title} {Statistical
  behavior in deterministic quantum systems with few degrees of freedom},}\
  }\href {\doibase 10.1103/PhysRevLett.54.1879} {\bibfield  {journal} {\bibinfo
   {journal} {Phys. Rev. Lett.}\ }\textbf {\bibinfo {volume} {54}},\ \bibinfo
  {pages} {1879--1882} (\bibinfo {year} {1985})}\BibitemShut {NoStop}%
\bibitem [{\citenamefont {Landau}\ and\ \citenamefont
  {Lifshitz}(1980)}]{Landau_Lifshitz}%
  \BibitemOpen
  \bibfield  {author} {\bibinfo {author} {\bibfnamefont {L.D.}\ \bibnamefont
  {Landau}}\ and\ \bibinfo {author} {\bibfnamefont {E.M.}\ \bibnamefont
  {Lifshitz}},\ }\href
  {https://www.sciencedirect.com/book/9780080570464/statistical-physics} {\emph
  {\bibinfo {title} {Statistical Physics}}},\ \bibinfo {edition} {third
  edition}\ ed.\ (\bibinfo  {publisher} {Butterworth-Heinemann},\ \bibinfo
  {address} {Oxford},\ \bibinfo {year} {1980})\BibitemShut {NoStop}%
\bibitem [{\citenamefont {Basko}\ \emph {et~al.}(2006)\citenamefont {Basko},
  \citenamefont {Aleiner},\ and\ \citenamefont {Altshuler}}]{Basko2006}%
  \BibitemOpen
  \bibfield  {author} {\bibinfo {author} {\bibfnamefont {D.~M.}\ \bibnamefont
  {Basko}}, \bibinfo {author} {\bibfnamefont {I.~L.}\ \bibnamefont {Aleiner}},
  \ and\ \bibinfo {author} {\bibfnamefont {B.~L.}\ \bibnamefont {Altshuler}},\
  }\bibfield  {title} {\enquote {\bibinfo {title} {Metal-insulator transition
  in a weakly interacting many-electron system with localized single-particle
  states},}\ }\href {\doibase 10.1016/j.aop.2005.11.014} {\bibfield  {journal}
  {\bibinfo  {journal} {Ann. Phys. (N.Y.)}\ }\textbf {\bibinfo {volume}
  {321}},\ \bibinfo {pages} {1126} (\bibinfo {year} {2006})}\BibitemShut
  {NoStop}%
\bibitem [{\citenamefont {Pal}\ and\ \citenamefont {Huse}(2010)}]{Pal2010}%
  \BibitemOpen
  \bibfield  {author} {\bibinfo {author} {\bibfnamefont {Arijeet}\ \bibnamefont
  {Pal}}\ and\ \bibinfo {author} {\bibfnamefont {David~A.}\ \bibnamefont
  {Huse}},\ }\bibfield  {title} {\enquote {\bibinfo {title} {Many-body
  localization phase transition},}\ }\href {\doibase
  10.1103/PhysRevB.82.174411} {\bibfield  {journal} {\bibinfo  {journal} {Phys.
  Rev. B}\ }\textbf {\bibinfo {volume} {82}},\ \bibinfo {pages} {174411}
  (\bibinfo {year} {2010})}\BibitemShut {NoStop}%
\bibitem [{\citenamefont {Oganesyan}\ and\ \citenamefont
  {Huse}(2007)}]{Oganesyan2007}%
  \BibitemOpen
  \bibfield  {author} {\bibinfo {author} {\bibfnamefont {Vadim}\ \bibnamefont
  {Oganesyan}}\ and\ \bibinfo {author} {\bibfnamefont {David~A.}\ \bibnamefont
  {Huse}},\ }\bibfield  {title} {\enquote {\bibinfo {title} {Localization of
  interacting fermions at high temperature},}\ }\href {\doibase
  10.1103/PhysRevB.75.155111} {\bibfield  {journal} {\bibinfo  {journal} {Phys.
  Rev. B}\ }\textbf {\bibinfo {volume} {75}},\ \bibinfo {pages} {155111}
  (\bibinfo {year} {2007})}\BibitemShut {NoStop}%
\bibitem [{\citenamefont {Nandkishore}\ and\ \citenamefont
  {Huse}(2015)}]{Nandkishore2015}%
  \BibitemOpen
  \bibfield  {author} {\bibinfo {author} {\bibfnamefont {Rahul}\ \bibnamefont
  {Nandkishore}}\ and\ \bibinfo {author} {\bibfnamefont {David~A.}\
  \bibnamefont {Huse}},\ }\bibfield  {title} {\enquote {\bibinfo {title}
  {Many-body localization and thermalization in quantum statistical
  mechanics},}\ }\href {\doibase 10.1146/annurev-conmatphys-031214-014726}
  {\bibfield  {journal} {\bibinfo  {journal} {Annu. Rev. Condens. Matter
  Phys.}\ }\textbf {\bibinfo {volume} {6}},\ \bibinfo {pages} {15–38}
  (\bibinfo {year} {2015})}\BibitemShut {NoStop}%
\bibitem [{\citenamefont {Abanin}\ \emph {et~al.}(2019)\citenamefont {Abanin},
  \citenamefont {Altman}, \citenamefont {Bloch},\ and\ \citenamefont
  {Serbyn}}]{Abanin2019_Review}%
  \BibitemOpen
  \bibfield  {author} {\bibinfo {author} {\bibfnamefont {Dmitry~A.}\
  \bibnamefont {Abanin}}, \bibinfo {author} {\bibfnamefont {Ehud}\ \bibnamefont
  {Altman}}, \bibinfo {author} {\bibfnamefont {Immanuel}\ \bibnamefont
  {Bloch}}, \ and\ \bibinfo {author} {\bibfnamefont {Maksym}\ \bibnamefont
  {Serbyn}},\ }\bibfield  {title} {\enquote {\bibinfo {title} {Colloquium:
  Many-body localization, thermalization, and entanglement},}\ }\href {\doibase
  10.1103/RevModPhys.91.021001} {\bibfield  {journal} {\bibinfo  {journal}
  {Rev. Mod. Phys.}\ }\textbf {\bibinfo {volume} {91}},\ \bibinfo {pages}
  {021001} (\bibinfo {year} {2019})}\BibitemShut {NoStop}%
\bibitem [{\citenamefont {Heller}(1984)}]{Heller1984}%
  \BibitemOpen
  \bibfield  {author} {\bibinfo {author} {\bibfnamefont {Eric~J.}\ \bibnamefont
  {Heller}},\ }\bibfield  {title} {\enquote {\bibinfo {title} {Bound-state
  eigenfunctions of classically chaotic {Hamiltonian} systems: Scars of
  periodic orbits},}\ }\href {\doibase 10.1103/PhysRevLett.53.1515} {\bibfield
  {journal} {\bibinfo  {journal} {Phys. Rev. Lett.}\ }\textbf {\bibinfo
  {volume} {53}},\ \bibinfo {pages} {1515} (\bibinfo {year}
  {1984})}\BibitemShut {NoStop}%
\bibitem [{\citenamefont {Vafek}\ \emph {et~al.}(2017)\citenamefont {Vafek},
  \citenamefont {Regnault},\ and\ \citenamefont {Bernevig}}]{Vafek2017}%
  \BibitemOpen
  \bibfield  {author} {\bibinfo {author} {\bibfnamefont {Oskar}\ \bibnamefont
  {Vafek}}, \bibinfo {author} {\bibfnamefont {Nicolas}\ \bibnamefont
  {Regnault}}, \ and\ \bibinfo {author} {\bibfnamefont {B.~Andrei}\
  \bibnamefont {Bernevig}},\ }\bibfield  {title} {\enquote {\bibinfo {title}
  {Entanglement of exact excited eigenstates of the {Hubbard} model in
  arbitrary dimension},}\ }\href {\doibase 10.21468/scipostphys.3.6.043}
  {\bibfield  {journal} {\bibinfo  {journal} {SciPost Phys.}\ }\textbf
  {\bibinfo {volume} {3}},\ \bibinfo {pages} {043} (\bibinfo {year}
  {2017})}\BibitemShut {NoStop}%
\bibitem [{\citenamefont {Shiraishi}\ and\ \citenamefont
  {Mori}(2017)}]{Shiraishi2017}%
  \BibitemOpen
  \bibfield  {author} {\bibinfo {author} {\bibfnamefont {Naoto}\ \bibnamefont
  {Shiraishi}}\ and\ \bibinfo {author} {\bibfnamefont {Takashi}\ \bibnamefont
  {Mori}},\ }\bibfield  {title} {\enquote {\bibinfo {title} {Systematic
  construction of counterexamples to the eigenstate thermalization
  hypothesis},}\ }\href {\doibase 10.1103/PhysRevLett.119.030601} {\bibfield
  {journal} {\bibinfo  {journal} {Phys. Rev. Lett.}\ }\textbf {\bibinfo
  {volume} {119}},\ \bibinfo {pages} {030601} (\bibinfo {year}
  {2017})}\BibitemShut {NoStop}%
\bibitem [{\citenamefont {Turner}\ \emph
  {et~al.}(2018{\natexlab{a}})\citenamefont {Turner}, \citenamefont
  {Michailidis}, \citenamefont {Abanin}, \citenamefont {Serbyn},\ and\
  \citenamefont {Papic}}]{Turner2018_np}%
  \BibitemOpen
  \bibfield  {author} {\bibinfo {author} {\bibfnamefont {C.~J.}\ \bibnamefont
  {Turner}}, \bibinfo {author} {\bibfnamefont {A.~A.}\ \bibnamefont
  {Michailidis}}, \bibinfo {author} {\bibfnamefont {D.~A.}\ \bibnamefont
  {Abanin}}, \bibinfo {author} {\bibfnamefont {M.}~\bibnamefont {Serbyn}}, \
  and\ \bibinfo {author} {\bibfnamefont {Z.}~\bibnamefont {Papic}},\ }\bibfield
   {title} {\enquote {\bibinfo {title} {Weak ergodicity breaking from quantum
  many-body scars},}\ }\href {\doibase 10.1038/s41567-018-0137-5} {\bibfield
  {journal} {\bibinfo  {journal} {Nat. Phys.}\ }\textbf {\bibinfo {volume}
  {14}},\ \bibinfo {pages} {745} (\bibinfo {year}
  {2018}{\natexlab{a}})}\BibitemShut {NoStop}%
\bibitem [{\citenamefont {Turner}\ \emph
  {et~al.}(2018{\natexlab{b}})\citenamefont {Turner}, \citenamefont
  {Michailidis}, \citenamefont {Abanin}, \citenamefont {Serbyn},\ and\
  \citenamefont {Papi\ifmmode~\acute{c}\else \'{c}\fi{}}}]{Turner2018_prb}%
  \BibitemOpen
  \bibfield  {author} {\bibinfo {author} {\bibfnamefont {C.~J.}\ \bibnamefont
  {Turner}}, \bibinfo {author} {\bibfnamefont {A.~A.}\ \bibnamefont
  {Michailidis}}, \bibinfo {author} {\bibfnamefont {D.~A.}\ \bibnamefont
  {Abanin}}, \bibinfo {author} {\bibfnamefont {M.}~\bibnamefont {Serbyn}}, \
  and\ \bibinfo {author} {\bibfnamefont {Z.}~\bibnamefont
  {Papi\ifmmode~\acute{c}\else \'{c}\fi{}}},\ }\bibfield  {title} {\enquote
  {\bibinfo {title} {Quantum scarred eigenstates in a {Rydberg} atom chain:
  Entanglement, breakdown of thermalization, and stability to perturbations},}\
  }\href {\doibase 10.1103/PhysRevB.98.155134} {\bibfield  {journal} {\bibinfo
  {journal} {Phys. Rev. B}\ }\textbf {\bibinfo {volume} {98}},\ \bibinfo
  {pages} {155134} (\bibinfo {year} {2018}{\natexlab{b}})}\BibitemShut
  {NoStop}%
\bibitem [{\citenamefont {Moudgalya}\ \emph
  {et~al.}(2018{\natexlab{a}})\citenamefont {Moudgalya}, \citenamefont
  {Rachel}, \citenamefont {Bernevig},\ and\ \citenamefont
  {Regnault}}]{Moudgalya2018_Exact}%
  \BibitemOpen
  \bibfield  {author} {\bibinfo {author} {\bibfnamefont {Sanjay}\ \bibnamefont
  {Moudgalya}}, \bibinfo {author} {\bibfnamefont {Stephan}\ \bibnamefont
  {Rachel}}, \bibinfo {author} {\bibfnamefont {B.~Andrei}\ \bibnamefont
  {Bernevig}}, \ and\ \bibinfo {author} {\bibfnamefont {Nicolas}\ \bibnamefont
  {Regnault}},\ }\bibfield  {title} {\enquote {\bibinfo {title} {Exact excited
  states of nonintegrable models},}\ }\href {\doibase
  10.1103/PhysRevB.98.235155} {\bibfield  {journal} {\bibinfo  {journal} {Phys.
  Rev. B}\ }\textbf {\bibinfo {volume} {98}},\ \bibinfo {pages} {235155}
  (\bibinfo {year} {2018}{\natexlab{a}})}\BibitemShut {NoStop}%
\bibitem [{\citenamefont {Moudgalya}\ \emph
  {et~al.}(2018{\natexlab{b}})\citenamefont {Moudgalya}, \citenamefont
  {Regnault},\ and\ \citenamefont {Bernevig}}]{Moudgalya2018_AKLT}%
  \BibitemOpen
  \bibfield  {author} {\bibinfo {author} {\bibfnamefont {Sanjay}\ \bibnamefont
  {Moudgalya}}, \bibinfo {author} {\bibfnamefont {Nicolas}\ \bibnamefont
  {Regnault}}, \ and\ \bibinfo {author} {\bibfnamefont {B.~Andrei}\
  \bibnamefont {Bernevig}},\ }\bibfield  {title} {\enquote {\bibinfo {title}
  {Entanglement of exact excited states of {Affleck-Kennedy-Lieb-Tasaki}
  models: Exact results, many-body scars, and violation of the strong
  eigenstate thermalization hypothesis},}\ }\href {\doibase
  10.1103/PhysRevB.98.235156} {\bibfield  {journal} {\bibinfo  {journal} {Phys.
  Rev. B}\ }\textbf {\bibinfo {volume} {98}},\ \bibinfo {pages} {235156}
  (\bibinfo {year} {2018}{\natexlab{b}})}\BibitemShut {NoStop}%
\bibitem [{\citenamefont {Lin}\ and\ \citenamefont
  {Motrunich}(2019)}]{Lin2019}%
  \BibitemOpen
  \bibfield  {author} {\bibinfo {author} {\bibfnamefont {Cheng-Ju}\
  \bibnamefont {Lin}}\ and\ \bibinfo {author} {\bibfnamefont {Olexei~I.}\
  \bibnamefont {Motrunich}},\ }\bibfield  {title} {\enquote {\bibinfo {title}
  {Exact quantum many-body scar states in the {Rydberg}-blockaded atom
  chain},}\ }\href {\doibase 10.1103/PhysRevLett.122.173401} {\bibfield
  {journal} {\bibinfo  {journal} {Phys. Rev. Lett.}\ }\textbf {\bibinfo
  {volume} {122}},\ \bibinfo {pages} {173401} (\bibinfo {year}
  {2019})}\BibitemShut {NoStop}%
\bibitem [{\citenamefont {Khemani}\ \emph {et~al.}(2019)\citenamefont
  {Khemani}, \citenamefont {Laumann},\ and\ \citenamefont
  {Chandran}}]{Khemani2019_RydbergIntegrable}%
  \BibitemOpen
  \bibfield  {author} {\bibinfo {author} {\bibfnamefont {Vedika}\ \bibnamefont
  {Khemani}}, \bibinfo {author} {\bibfnamefont {Chris~R.}\ \bibnamefont
  {Laumann}}, \ and\ \bibinfo {author} {\bibfnamefont {Anushya}\ \bibnamefont
  {Chandran}},\ }\bibfield  {title} {\enquote {\bibinfo {title} {Signatures of
  integrability in the dynamics of {Rydberg}-blockaded chains},}\ }\href
  {\doibase 10.1103/PhysRevB.99.161101} {\bibfield  {journal} {\bibinfo
  {journal} {Phys. Rev. B}\ }\textbf {\bibinfo {volume} {99}},\ \bibinfo
  {pages} {161101(R)} (\bibinfo {year} {2019})}\BibitemShut {NoStop}%
\bibitem [{\citenamefont {Bull}\ \emph {et~al.}(2019)\citenamefont {Bull},
  \citenamefont {Martin},\ and\ \citenamefont {Papi\ifmmode~\acute{c}\else
  \'{c}\fi{}}}]{Bull2019}%
  \BibitemOpen
  \bibfield  {author} {\bibinfo {author} {\bibfnamefont {Kieran}\ \bibnamefont
  {Bull}}, \bibinfo {author} {\bibfnamefont {Ivar}\ \bibnamefont {Martin}}, \
  and\ \bibinfo {author} {\bibfnamefont {Z.}~\bibnamefont
  {Papi\ifmmode~\acute{c}\else \'{c}\fi{}}},\ }\bibfield  {title} {\enquote
  {\bibinfo {title} {Systematic construction of scarred many-body dynamics in
  {1D} lattice models},}\ }\href {\doibase 10.1103/PhysRevLett.123.030601}
  {\bibfield  {journal} {\bibinfo  {journal} {Phys. Rev. Lett.}\ }\textbf
  {\bibinfo {volume} {123}},\ \bibinfo {pages} {030601} (\bibinfo {year}
  {2019})}\BibitemShut {NoStop}%
\bibitem [{\citenamefont {Schecter}\ and\ \citenamefont
  {Iadecola}(2019)}]{Schecter2019}%
  \BibitemOpen
  \bibfield  {author} {\bibinfo {author} {\bibfnamefont {Michael}\ \bibnamefont
  {Schecter}}\ and\ \bibinfo {author} {\bibfnamefont {Thomas}\ \bibnamefont
  {Iadecola}},\ }\bibfield  {title} {\enquote {\bibinfo {title} {Weak
  ergodicity breaking and quantum many-body scars in spin-1 {{\textit{XY}}}
  magnets},}\ }\href {\doibase 10.1103/PhysRevLett.123.147201} {\bibfield
  {journal} {\bibinfo  {journal} {Phys. Rev. Lett.}\ }\textbf {\bibinfo
  {volume} {123}},\ \bibinfo {pages} {147201} (\bibinfo {year}
  {2019})}\BibitemShut {NoStop}%
\bibitem [{\citenamefont {Ok}\ \emph {et~al.}(2019)\citenamefont {Ok},
  \citenamefont {Choo}, \citenamefont {Mudry}, \citenamefont {Castelnovo},
  \citenamefont {Chamon},\ and\ \citenamefont {Neupert}}]{Ok2019}%
  \BibitemOpen
  \bibfield  {author} {\bibinfo {author} {\bibfnamefont {Seulgi}\ \bibnamefont
  {Ok}}, \bibinfo {author} {\bibfnamefont {Kenny}\ \bibnamefont {Choo}},
  \bibinfo {author} {\bibfnamefont {Christopher}\ \bibnamefont {Mudry}},
  \bibinfo {author} {\bibfnamefont {Claudio}\ \bibnamefont {Castelnovo}},
  \bibinfo {author} {\bibfnamefont {Claudio}\ \bibnamefont {Chamon}}, \ and\
  \bibinfo {author} {\bibfnamefont {Titus}\ \bibnamefont {Neupert}},\
  }\bibfield  {title} {\enquote {\bibinfo {title} {Topological many-body scar
  states in dimensions one, two, and three},}\ }\href {\doibase
  10.1103/PhysRevResearch.1.033144} {\bibfield  {journal} {\bibinfo  {journal}
  {Phys. Rev. Research}\ }\textbf {\bibinfo {volume} {1}},\ \bibinfo {pages}
  {033144} (\bibinfo {year} {2019})}\BibitemShut {NoStop}%
\bibitem [{\citenamefont {Lee}\ \emph {et~al.}(2020)\citenamefont {Lee},
  \citenamefont {Melendrez}, \citenamefont {Pal},\ and\ \citenamefont
  {Changlani}}]{Lee_PRBR2020}%
  \BibitemOpen
  \bibfield  {author} {\bibinfo {author} {\bibfnamefont {Kyungmin}\
  \bibnamefont {Lee}}, \bibinfo {author} {\bibfnamefont {Ronald}\ \bibnamefont
  {Melendrez}}, \bibinfo {author} {\bibfnamefont {Arijeet}\ \bibnamefont
  {Pal}}, \ and\ \bibinfo {author} {\bibfnamefont {Hitesh~J.}\ \bibnamefont
  {Changlani}},\ }\bibfield  {title} {\enquote {\bibinfo {title} {Exact
  three-colored quantum scars from geometric frustration},}\ }\href {\doibase
  10.1103/PhysRevB.101.241111} {\bibfield  {journal} {\bibinfo  {journal}
  {Phys. Rev. B}\ }\textbf {\bibinfo {volume} {101}},\ \bibinfo {pages}
  {241111} (\bibinfo {year} {2020})}\BibitemShut {NoStop}%
\bibitem [{\citenamefont {Lee}\ \emph {et~al.}(2021)\citenamefont {Lee},
  \citenamefont {Pal},\ and\ \citenamefont {Changlani}}]{Lee_Pal_Changlani}%
  \BibitemOpen
  \bibfield  {author} {\bibinfo {author} {\bibfnamefont {Kyungmin}\
  \bibnamefont {Lee}}, \bibinfo {author} {\bibfnamefont {Arijeet}\ \bibnamefont
  {Pal}}, \ and\ \bibinfo {author} {\bibfnamefont {Hitesh~J.}\ \bibnamefont
  {Changlani}},\ }\bibfield  {title} {\enquote {\bibinfo {title}
  {Frustration-induced emergent hilbert space fragmentation},}\ }\href
  {\doibase 10.1103/PhysRevB.103.235133} {\bibfield  {journal} {\bibinfo
  {journal} {Phys. Rev. B}\ }\textbf {\bibinfo {volume} {103}},\ \bibinfo
  {pages} {235133} (\bibinfo {year} {2021})}\BibitemShut {NoStop}%
\bibitem [{\citenamefont {McClarty}\ \emph {et~al.}(2020)\citenamefont
  {McClarty}, \citenamefont {Haque}, \citenamefont {Sen},\ and\ \citenamefont
  {Richter}}]{McClartyscar}%
  \BibitemOpen
  \bibfield  {author} {\bibinfo {author} {\bibfnamefont {Paul~A.}\ \bibnamefont
  {McClarty}}, \bibinfo {author} {\bibfnamefont {Masudul}\ \bibnamefont
  {Haque}}, \bibinfo {author} {\bibfnamefont {Arnab}\ \bibnamefont {Sen}}, \
  and\ \bibinfo {author} {\bibfnamefont {Johannes}\ \bibnamefont {Richter}},\
  }\bibfield  {title} {\enquote {\bibinfo {title} {Disorder-free localization
  and many-body quantum scars from magnetic frustration},}\ }\href {\doibase
  10.1103/PhysRevB.102.224303} {\bibfield  {journal} {\bibinfo  {journal}
  {Phys. Rev. B}\ }\textbf {\bibinfo {volume} {102}},\ \bibinfo {pages}
  {224303} (\bibinfo {year} {2020})}\BibitemShut {NoStop}%
\bibitem [{\citenamefont {Wildeboer}\ \emph {et~al.}(2021)\citenamefont
  {Wildeboer}, \citenamefont {Seidel}, \citenamefont {Srivatsa}, \citenamefont
  {Nielsen},\ and\ \citenamefont {Erten}}]{Wildeboer_scar}%
  \BibitemOpen
  \bibfield  {author} {\bibinfo {author} {\bibfnamefont {Julia}\ \bibnamefont
  {Wildeboer}}, \bibinfo {author} {\bibfnamefont {Alexander}\ \bibnamefont
  {Seidel}}, \bibinfo {author} {\bibfnamefont {N.~S.}\ \bibnamefont
  {Srivatsa}}, \bibinfo {author} {\bibfnamefont {Anne E.~B.}\ \bibnamefont
  {Nielsen}}, \ and\ \bibinfo {author} {\bibfnamefont {Onur}\ \bibnamefont
  {Erten}},\ }\bibfield  {title} {\enquote {\bibinfo {title} {Topological
  quantum many-body scars in quantum dimer models on the kagome lattice},}\
  }\href {\doibase 10.1103/PhysRevB.104.L121103} {\bibfield  {journal}
  {\bibinfo  {journal} {Phys. Rev. B}\ }\textbf {\bibinfo {volume} {104}},\
  \bibinfo {pages} {L121103} (\bibinfo {year} {2021})}\BibitemShut {NoStop}%
\bibitem [{\citenamefont {Pakrouski}\ \emph {et~al.}(2020)\citenamefont
  {Pakrouski}, \citenamefont {Pallegar}, \citenamefont {Popov},\ and\
  \citenamefont {Klebanov}}]{Pakrouski_scar}%
  \BibitemOpen
  \bibfield  {author} {\bibinfo {author} {\bibfnamefont {K.}~\bibnamefont
  {Pakrouski}}, \bibinfo {author} {\bibfnamefont {P.~N.}\ \bibnamefont
  {Pallegar}}, \bibinfo {author} {\bibfnamefont {F.~K.}\ \bibnamefont {Popov}},
  \ and\ \bibinfo {author} {\bibfnamefont {I.~R.}\ \bibnamefont {Klebanov}},\
  }\bibfield  {title} {\enquote {\bibinfo {title} {Many-body scars as a group
  invariant sector of hilbert space},}\ }\href {\doibase
  10.1103/PhysRevLett.125.230602} {\bibfield  {journal} {\bibinfo  {journal}
  {Phys. Rev. Lett.}\ }\textbf {\bibinfo {volume} {125}},\ \bibinfo {pages}
  {230602} (\bibinfo {year} {2020})}\BibitemShut {NoStop}%
\bibitem [{\citenamefont {van Voorden}\ \emph {et~al.}(2020)\citenamefont {van
  Voorden}, \citenamefont {Min\'a\ifmmode~\check{r}\else \v{r}\fi{}},\ and\
  \citenamefont {Schoutens}}]{Voorden_scar}%
  \BibitemOpen
  \bibfield  {author} {\bibinfo {author} {\bibfnamefont {Bart}\ \bibnamefont
  {van Voorden}}, \bibinfo {author} {\bibfnamefont {Ji\ifmmode
  \check{r}\else~\v{r}\fi{}\'{\i}}\ \bibnamefont {Min\'a\ifmmode~\check{r}\else
  \v{r}\fi{}}}, \ and\ \bibinfo {author} {\bibfnamefont {Kareljan}\
  \bibnamefont {Schoutens}},\ }\bibfield  {title} {\enquote {\bibinfo {title}
  {Quantum many-body scars in transverse field ising ladders and beyond},}\
  }\href {\doibase 10.1103/PhysRevB.101.220305} {\bibfield  {journal} {\bibinfo
   {journal} {Phys. Rev. B}\ }\textbf {\bibinfo {volume} {101}},\ \bibinfo
  {pages} {220305} (\bibinfo {year} {2020})}\BibitemShut {NoStop}%
\bibitem [{\citenamefont {You}\ \emph {et~al.}(2022)\citenamefont {You},
  \citenamefont {Zhao}, \citenamefont {Ren}, \citenamefont {Sun}, \citenamefont
  {Li},\ and\ \citenamefont {Ole\ifmmode~\acute{s}\else
  \'{s}\fi{}}}]{Spin1Kitaevscars}%
  \BibitemOpen
  \bibfield  {author} {\bibinfo {author} {\bibfnamefont {Wen-Long}\
  \bibnamefont {You}}, \bibinfo {author} {\bibfnamefont {Zhuan}\ \bibnamefont
  {Zhao}}, \bibinfo {author} {\bibfnamefont {Jie}\ \bibnamefont {Ren}},
  \bibinfo {author} {\bibfnamefont {Gaoyong}\ \bibnamefont {Sun}}, \bibinfo
  {author} {\bibfnamefont {Liangsheng}\ \bibnamefont {Li}}, \ and\ \bibinfo
  {author} {\bibfnamefont {Andrzej~M.}\ \bibnamefont
  {Ole\ifmmode~\acute{s}\else \'{s}\fi{}}},\ }\bibfield  {title} {\enquote
  {\bibinfo {title} {Quantum many-body scars in spin-1 kitaev chains},}\ }\href
  {\doibase 10.1103/PhysRevResearch.4.013103} {\bibfield  {journal} {\bibinfo
  {journal} {Phys. Rev. Research}\ }\textbf {\bibinfo {volume} {4}},\ \bibinfo
  {pages} {013103} (\bibinfo {year} {2022})}\BibitemShut {NoStop}%
\bibitem [{\citenamefont {Chertkov}\ and\ \citenamefont
  {Clark}(2021)}]{Chertkovscar}%
  \BibitemOpen
  \bibfield  {author} {\bibinfo {author} {\bibfnamefont {Eli}\ \bibnamefont
  {Chertkov}}\ and\ \bibinfo {author} {\bibfnamefont {Bryan~K.}\ \bibnamefont
  {Clark}},\ }\bibfield  {title} {\enquote {\bibinfo {title} {Motif magnetim
  and quantum many-body scars},}\ }\href {\doibase 10.1103/PhysRevB.104.104410}
  {\bibfield  {journal} {\bibinfo  {journal} {Phys. Rev. B}\ }\textbf {\bibinfo
  {volume} {104}},\ \bibinfo {pages} {104410} (\bibinfo {year}
  {2021})}\BibitemShut {NoStop}%
\bibitem [{\citenamefont {Kuno}\ \emph {et~al.}(2020)\citenamefont {Kuno},
  \citenamefont {Mizoguchi},\ and\ \citenamefont {Hatsugai}}]{Hatsugai_scar}%
  \BibitemOpen
  \bibfield  {author} {\bibinfo {author} {\bibfnamefont {Yoshihito}\
  \bibnamefont {Kuno}}, \bibinfo {author} {\bibfnamefont {Tomonari}\
  \bibnamefont {Mizoguchi}}, \ and\ \bibinfo {author} {\bibfnamefont
  {Yasuhiro}\ \bibnamefont {Hatsugai}},\ }\bibfield  {title} {\enquote
  {\bibinfo {title} {Flat band quantum scar},}\ }\href {\doibase
  10.1103/PhysRevB.102.241115} {\bibfield  {journal} {\bibinfo  {journal}
  {Phys. Rev. B}\ }\textbf {\bibinfo {volume} {102}},\ \bibinfo {pages}
  {241115} (\bibinfo {year} {2020})}\BibitemShut {NoStop}%
\bibitem [{\citenamefont {Ma}\ \emph {et~al.}(2022{\natexlab{a}})\citenamefont
  {Ma}, \citenamefont {Zhang},\ and\ \citenamefont {Song}}]{Song_2022}%
  \BibitemOpen
  \bibfield  {author} {\bibinfo {author} {\bibfnamefont {E.~S.}\ \bibnamefont
  {Ma}}, \bibinfo {author} {\bibfnamefont {K.~L.}\ \bibnamefont {Zhang}}, \
  and\ \bibinfo {author} {\bibfnamefont {Z.}~\bibnamefont {Song}},\ }\bibfield
  {title} {\enquote {\bibinfo {title} {Steady helix states in a resonant xxz
  heisenberg model with dzyaloshinskii-moriya interaction},}\ }\href {\doibase
  10.1103/PhysRevB.106.245122} {\bibfield  {journal} {\bibinfo  {journal}
  {Phys. Rev. B}\ }\textbf {\bibinfo {volume} {106}},\ \bibinfo {pages}
  {245122} (\bibinfo {year} {2022}{\natexlab{a}})}\BibitemShut {NoStop}%
\bibitem [{\citenamefont {Sharma}\ \emph {et~al.}(2022)\citenamefont {Sharma},
  \citenamefont {Lee},\ and\ \citenamefont {Changlani}}]{Sharma_spin1}%
  \BibitemOpen
  \bibfield  {author} {\bibinfo {author} {\bibfnamefont {Prakash}\ \bibnamefont
  {Sharma}}, \bibinfo {author} {\bibfnamefont {Kyungmin}\ \bibnamefont {Lee}},
  \ and\ \bibinfo {author} {\bibfnamefont {Hitesh~J.}\ \bibnamefont
  {Changlani}},\ }\bibfield  {title} {\enquote {\bibinfo {title} {Multimagnon
  dynamics and thermalization in the $s=1$ easy-axis ferromagnetic chain},}\
  }\href {\doibase 10.1103/PhysRevB.105.054413} {\bibfield  {journal} {\bibinfo
   {journal} {Phys. Rev. B}\ }\textbf {\bibinfo {volume} {105}},\ \bibinfo
  {pages} {054413} (\bibinfo {year} {2022})}\BibitemShut {NoStop}%
\bibitem [{\citenamefont {Pilatowsky-Cameo}\ \emph {et~al.}(2021)\citenamefont
  {Pilatowsky-Cameo}, \citenamefont {Villase{\~{n}}or}, \citenamefont
  {Bastarrachea-Magnani}, \citenamefont {Lerma-Hern{\'a}ndez}, \citenamefont
  {Santos},\ and\ \citenamefont {Hirsch}}]{Hirsch_scar}%
  \BibitemOpen
  \bibfield  {author} {\bibinfo {author} {\bibfnamefont {Sa{\'u}l}\
  \bibnamefont {Pilatowsky-Cameo}}, \bibinfo {author} {\bibfnamefont {David}\
  \bibnamefont {Villase{\~{n}}or}}, \bibinfo {author} {\bibfnamefont
  {Miguel~A.}\ \bibnamefont {Bastarrachea-Magnani}}, \bibinfo {author}
  {\bibfnamefont {Sergio}\ \bibnamefont {Lerma-Hern{\'a}ndez}}, \bibinfo
  {author} {\bibfnamefont {Lea~F.}\ \bibnamefont {Santos}}, \ and\ \bibinfo
  {author} {\bibfnamefont {Jorge~G.}\ \bibnamefont {Hirsch}},\ }\bibfield
  {title} {\enquote {\bibinfo {title} {Ubiquitous quantum scarring does not
  prevent ergodicity},}\ }\href {\doibase 10.1038/s41467-021-21123-5}
  {\bibfield  {journal} {\bibinfo  {journal} {Nature Communications}\ }\textbf
  {\bibinfo {volume} {12}},\ \bibinfo {pages} {852} (\bibinfo {year}
  {2021})}\BibitemShut {NoStop}%
\bibitem [{\citenamefont {Serbyn}\ \emph {et~al.}(2020)\citenamefont {Serbyn},
  \citenamefont {Abanin},\ and\ \citenamefont {Papić}}]{Serbyn_review}%
  \BibitemOpen
  \bibfield  {author} {\bibinfo {author} {\bibfnamefont {Maksym}\ \bibnamefont
  {Serbyn}}, \bibinfo {author} {\bibfnamefont {Dmitry~A.}\ \bibnamefont
  {Abanin}}, \ and\ \bibinfo {author} {\bibfnamefont {Zlatko}\ \bibnamefont
  {Papić}},\ }\href@noop {} {\enquote {\bibinfo {title} {Quantum many-body
  scars and weak breaking of ergodicity},}\ } (\bibinfo {year} {2020}),\
  \Eprint {http://arxiv.org/abs/2011.09486} {arXiv:2011.09486 [quant-ph]}
  \BibitemShut {NoStop}%
\bibitem [{\citenamefont {Chandran}\ \emph {et~al.}(2022)\citenamefont
  {Chandran}, \citenamefont {Iadecola}, \citenamefont {Khemani},\ and\
  \citenamefont {Moessner}}]{Chandran_review}%
  \BibitemOpen
  \bibfield  {author} {\bibinfo {author} {\bibfnamefont {Anushya}\ \bibnamefont
  {Chandran}}, \bibinfo {author} {\bibfnamefont {Thomas}\ \bibnamefont
  {Iadecola}}, \bibinfo {author} {\bibfnamefont {Vedika}\ \bibnamefont
  {Khemani}}, \ and\ \bibinfo {author} {\bibfnamefont {Roderich}\ \bibnamefont
  {Moessner}},\ }\href {\doibase 10.48550/ARXIV.2206.11528} {\enquote {\bibinfo
  {title} {Quantum many-body scars: A quasiparticle perspective},}\ } (\bibinfo
  {year} {2022})\BibitemShut {NoStop}%
\bibitem [{\citenamefont {Dooley}(2021)}]{Dooley}%
  \BibitemOpen
  \bibfield  {author} {\bibinfo {author} {\bibfnamefont {Shane}\ \bibnamefont
  {Dooley}},\ }\bibfield  {title} {\enquote {\bibinfo {title} {Robust quantum
  sensing in strongly interacting systems with many-body scars},}\ }\href
  {\doibase 10.1103/PRXQuantum.2.020330} {\bibfield  {journal} {\bibinfo
  {journal} {PRX Quantum}\ }\textbf {\bibinfo {volume} {2}},\ \bibinfo {pages}
  {020330} (\bibinfo {year} {2021})}\BibitemShut {NoStop}%
\bibitem [{\citenamefont {Ma}\ \emph {et~al.}(2022{\natexlab{b}})\citenamefont
  {Ma}, \citenamefont {Volya},\ and\ \citenamefont {Yang}}]{Ken_2022}%
  \BibitemOpen
  \bibfield  {author} {\bibinfo {author} {\bibfnamefont {Ken K.~W.}\
  \bibnamefont {Ma}}, \bibinfo {author} {\bibfnamefont {A.}~\bibnamefont
  {Volya}}, \ and\ \bibinfo {author} {\bibfnamefont {Kun}\ \bibnamefont
  {Yang}},\ }\bibfield  {title} {\enquote {\bibinfo {title} {Eigenstate
  thermalization and disappearance of quantum many-body scar states in weakly
  interacting fermion systems},}\ }\href {\doibase 10.1103/PhysRevB.106.214313}
  {\bibfield  {journal} {\bibinfo  {journal} {Phys. Rev. B}\ }\textbf {\bibinfo
  {volume} {106}},\ \bibinfo {pages} {214313} (\bibinfo {year}
  {2022}{\natexlab{b}})}\BibitemShut {NoStop}%
\bibitem [{\citenamefont {Sala}\ \emph {et~al.}(2020)\citenamefont {Sala},
  \citenamefont {Rakovszky}, \citenamefont {Verresen}, \citenamefont {Knap},\
  and\ \citenamefont {Pollmann}}]{sala2019ergodicity}%
  \BibitemOpen
  \bibfield  {author} {\bibinfo {author} {\bibfnamefont {Pablo}\ \bibnamefont
  {Sala}}, \bibinfo {author} {\bibfnamefont {Tibor}\ \bibnamefont {Rakovszky}},
  \bibinfo {author} {\bibfnamefont {Ruben}\ \bibnamefont {Verresen}}, \bibinfo
  {author} {\bibfnamefont {Michael}\ \bibnamefont {Knap}}, \ and\ \bibinfo
  {author} {\bibfnamefont {Frank}\ \bibnamefont {Pollmann}},\ }\bibfield
  {title} {\enquote {\bibinfo {title} {Ergodicity breaking arising from
  {Hilbert} space fragmentation in dipole-conserving {Hamiltonians}},}\ }\href
  {\doibase 10.1103/PhysRevX.10.011047} {\bibfield  {journal} {\bibinfo
  {journal} {Phys. Rev. X}\ }\textbf {\bibinfo {volume} {10}},\ \bibinfo
  {pages} {011047} (\bibinfo {year} {2020})}\BibitemShut {NoStop}%
\bibitem [{\citenamefont {Khemani}\ \emph {et~al.}(2020)\citenamefont
  {Khemani}, \citenamefont {Hermele},\ and\ \citenamefont
  {Nandkishore}}]{khemani2019local}%
  \BibitemOpen
  \bibfield  {author} {\bibinfo {author} {\bibfnamefont {Vedika}\ \bibnamefont
  {Khemani}}, \bibinfo {author} {\bibfnamefont {Michael}\ \bibnamefont
  {Hermele}}, \ and\ \bibinfo {author} {\bibfnamefont {Rahul}\ \bibnamefont
  {Nandkishore}},\ }\bibfield  {title} {\enquote {\bibinfo {title}
  {Localization from {Hilbert} space shattering: From theory to physical
  realizations},}\ }\href {\doibase 10.1103/PhysRevB.101.174204} {\bibfield
  {journal} {\bibinfo  {journal} {Phys. Rev. B}\ }\textbf {\bibinfo {volume}
  {101}},\ \bibinfo {pages} {174204} (\bibinfo {year} {2020})}\BibitemShut
  {NoStop}%
\bibitem [{\citenamefont {Khudorozhkov}\ \emph {et~al.}(2022)\citenamefont
  {Khudorozhkov}, \citenamefont {Tiwari}, \citenamefont {Chamon},\ and\
  \citenamefont {Neupert}}]{Neupert_HSF}%
  \BibitemOpen
  \bibfield  {author} {\bibinfo {author} {\bibfnamefont {Alexey}\ \bibnamefont
  {Khudorozhkov}}, \bibinfo {author} {\bibfnamefont {Apoorv}\ \bibnamefont
  {Tiwari}}, \bibinfo {author} {\bibfnamefont {Claudio}\ \bibnamefont
  {Chamon}}, \ and\ \bibinfo {author} {\bibfnamefont {Titus}\ \bibnamefont
  {Neupert}},\ }\bibfield  {title} {\enquote {\bibinfo {title} {{Hilbert space
  fragmentation in a 2D quantum spin system with subsystem symmetries}},}\
  }\href {\doibase 10.21468/SciPostPhys.13.4.098} {\bibfield  {journal}
  {\bibinfo  {journal} {SciPost Phys.}\ }\textbf {\bibinfo {volume} {13}},\
  \bibinfo {pages} {098} (\bibinfo {year} {2022})}\BibitemShut {NoStop}%
\bibitem [{\citenamefont {Moudgalya}\ and\ \citenamefont
  {Motrunich}(2022)}]{Moudgalya_HSF}%
  \BibitemOpen
  \bibfield  {author} {\bibinfo {author} {\bibfnamefont {Sanjay}\ \bibnamefont
  {Moudgalya}}\ and\ \bibinfo {author} {\bibfnamefont {Olexei~I.}\ \bibnamefont
  {Motrunich}},\ }\bibfield  {title} {\enquote {\bibinfo {title} {Hilbert space
  fragmentation and commutant algebras},}\ }\href {\doibase
  10.1103/PhysRevX.12.011050} {\bibfield  {journal} {\bibinfo  {journal} {Phys.
  Rev. X}\ }\textbf {\bibinfo {volume} {12}},\ \bibinfo {pages} {011050}
  (\bibinfo {year} {2022})}\BibitemShut {NoStop}%
\bibitem [{\citenamefont {Mukherjee}\ \emph
  {et~al.}(2020{\natexlab{a}})\citenamefont {Mukherjee}, \citenamefont {Nandy},
  \citenamefont {Sen}, \citenamefont {Sen},\ and\ \citenamefont
  {Sengupta}}]{Mukherjee_2020}%
  \BibitemOpen
  \bibfield  {author} {\bibinfo {author} {\bibfnamefont {Bhaskar}\ \bibnamefont
  {Mukherjee}}, \bibinfo {author} {\bibfnamefont {Sourav}\ \bibnamefont
  {Nandy}}, \bibinfo {author} {\bibfnamefont {Arnab}\ \bibnamefont {Sen}},
  \bibinfo {author} {\bibfnamefont {Diptiman}\ \bibnamefont {Sen}}, \ and\
  \bibinfo {author} {\bibfnamefont {K.}~\bibnamefont {Sengupta}},\ }\bibfield
  {title} {\enquote {\bibinfo {title} {Collapse and revival of quantum
  many-body scars via floquet engineering},}\ }\href {\doibase
  10.1103/PhysRevB.101.245107} {\bibfield  {journal} {\bibinfo  {journal}
  {Phys. Rev. B}\ }\textbf {\bibinfo {volume} {101}},\ \bibinfo {pages}
  {245107} (\bibinfo {year} {2020}{\natexlab{a}})}\BibitemShut {NoStop}%
\bibitem [{\citenamefont {Zhao}\ \emph {et~al.}(2020)\citenamefont {Zhao},
  \citenamefont {Vovrosh}, \citenamefont {Mintert},\ and\ \citenamefont
  {Knolle}}]{zhao}%
  \BibitemOpen
  \bibfield  {author} {\bibinfo {author} {\bibfnamefont {Hongzheng}\
  \bibnamefont {Zhao}}, \bibinfo {author} {\bibfnamefont {Joseph}\ \bibnamefont
  {Vovrosh}}, \bibinfo {author} {\bibfnamefont {Florian}\ \bibnamefont
  {Mintert}}, \ and\ \bibinfo {author} {\bibfnamefont {Johannes}\ \bibnamefont
  {Knolle}},\ }\bibfield  {title} {\enquote {\bibinfo {title} {Quantum
  many-body scars in optical lattices},}\ }\href {\doibase
  10.1103/PhysRevLett.124.160604} {\bibfield  {journal} {\bibinfo  {journal}
  {Phys. Rev. Lett.}\ }\textbf {\bibinfo {volume} {124}},\ \bibinfo {pages}
  {160604} (\bibinfo {year} {2020})}\BibitemShut {NoStop}%
\bibitem [{\citenamefont {Bluvstein}\ \emph {et~al.}(2021)\citenamefont
  {Bluvstein}, \citenamefont {Omran}, \citenamefont {Levine}, \citenamefont
  {Keesling}, \citenamefont {Semeghini}, \citenamefont {Ebadi}, \citenamefont
  {Wang}, \citenamefont {Michailidis}, \citenamefont {Maskara}, \citenamefont
  {Ho},\ and\ \citenamefont {et~al.}}]{Bluvstein_2021}%
  \BibitemOpen
  \bibfield  {author} {\bibinfo {author} {\bibfnamefont {D.}~\bibnamefont
  {Bluvstein}}, \bibinfo {author} {\bibfnamefont {A.}~\bibnamefont {Omran}},
  \bibinfo {author} {\bibfnamefont {H.}~\bibnamefont {Levine}}, \bibinfo
  {author} {\bibfnamefont {A.}~\bibnamefont {Keesling}}, \bibinfo {author}
  {\bibfnamefont {G.}~\bibnamefont {Semeghini}}, \bibinfo {author}
  {\bibfnamefont {S.}~\bibnamefont {Ebadi}}, \bibinfo {author} {\bibfnamefont
  {T.~T.}\ \bibnamefont {Wang}}, \bibinfo {author} {\bibfnamefont {A.~A.}\
  \bibnamefont {Michailidis}}, \bibinfo {author} {\bibfnamefont
  {N.}~\bibnamefont {Maskara}}, \bibinfo {author} {\bibfnamefont {W.~W.}\
  \bibnamefont {Ho}}, \ and\ \bibinfo {author} {\bibnamefont {et~al.}},\
  }\bibfield  {title} {\enquote {\bibinfo {title} {Controlling quantum
  many-body dynamics in driven rydberg atom arrays},}\ }\href {\doibase
  10.1126/science.abg2530} {\bibfield  {journal} {\bibinfo  {journal}
  {Science}\ }\textbf {\bibinfo {volume} {371}},\ \bibinfo {pages}
  {1355–1359} (\bibinfo {year} {2021})}\BibitemShut {NoStop}%
\bibitem [{\citenamefont {Mukherjee}\ \emph
  {et~al.}(2020{\natexlab{b}})\citenamefont {Mukherjee}, \citenamefont {Sen},
  \citenamefont {Sen},\ and\ \citenamefont {Sengupta}}]{freezing}%
  \BibitemOpen
  \bibfield  {author} {\bibinfo {author} {\bibfnamefont {Bhaskar}\ \bibnamefont
  {Mukherjee}}, \bibinfo {author} {\bibfnamefont {Arnab}\ \bibnamefont {Sen}},
  \bibinfo {author} {\bibfnamefont {Diptiman}\ \bibnamefont {Sen}}, \ and\
  \bibinfo {author} {\bibfnamefont {K.}~\bibnamefont {Sengupta}},\ }\bibfield
  {title} {\enquote {\bibinfo {title} {Dynamics of the vacuum state in a
  periodically driven rydberg chain},}\ }\href {\doibase
  10.1103/PhysRevB.102.075123} {\bibfield  {journal} {\bibinfo  {journal}
  {Phys. Rev. B}\ }\textbf {\bibinfo {volume} {102}},\ \bibinfo {pages}
  {075123} (\bibinfo {year} {2020}{\natexlab{b}})}\BibitemShut {NoStop}%
\bibitem [{\citenamefont {Haldar}\ \emph {et~al.}(2021)\citenamefont {Haldar},
  \citenamefont {Sen}, \citenamefont {Moessner},\ and\ \citenamefont
  {Das}}]{AsmiPRXResonant}%
  \BibitemOpen
  \bibfield  {author} {\bibinfo {author} {\bibfnamefont {Asmi}\ \bibnamefont
  {Haldar}}, \bibinfo {author} {\bibfnamefont {Diptiman}\ \bibnamefont {Sen}},
  \bibinfo {author} {\bibfnamefont {Roderich}\ \bibnamefont {Moessner}}, \ and\
  \bibinfo {author} {\bibfnamefont {Arnab}\ \bibnamefont {Das}},\ }\bibfield
  {title} {\enquote {\bibinfo {title} {Dynamical freezing and scar points in
  strongly driven floquet matter: Resonance vs emergent conservation laws},}\
  }\href {\doibase 10.1103/PhysRevX.11.021008} {\bibfield  {journal} {\bibinfo
  {journal} {Phys. Rev. X}\ }\textbf {\bibinfo {volume} {11}},\ \bibinfo
  {pages} {021008} (\bibinfo {year} {2021})}\BibitemShut {NoStop}%
\bibitem [{\citenamefont {Thuberg}\ \emph {et~al.}(2016)\citenamefont
  {Thuberg}, \citenamefont {Reyes},\ and\ \citenamefont {Eggert}}]{Thuberg}%
  \BibitemOpen
  \bibfield  {author} {\bibinfo {author} {\bibfnamefont {Daniel}\ \bibnamefont
  {Thuberg}}, \bibinfo {author} {\bibfnamefont {Sebasti\'an~A.}\ \bibnamefont
  {Reyes}}, \ and\ \bibinfo {author} {\bibfnamefont {Sebastian}\ \bibnamefont
  {Eggert}},\ }\bibfield  {title} {\enquote {\bibinfo {title} {Quantum
  resonance catastrophe for conductance through a periodically driven
  barrier},}\ }\href {\doibase 10.1103/PhysRevB.93.180301} {\bibfield
  {journal} {\bibinfo  {journal} {Phys. Rev. B}\ }\textbf {\bibinfo {volume}
  {93}},\ \bibinfo {pages} {180301} (\bibinfo {year} {2016})}\BibitemShut
  {NoStop}%
\bibitem [{\citenamefont {Agarwala}\ and\ \citenamefont {Sen}(2017)}]{adhip}%
  \BibitemOpen
  \bibfield  {author} {\bibinfo {author} {\bibfnamefont {Adhip}\ \bibnamefont
  {Agarwala}}\ and\ \bibinfo {author} {\bibfnamefont {Diptiman}\ \bibnamefont
  {Sen}},\ }\bibfield  {title} {\enquote {\bibinfo {title} {Effects of local
  periodic driving on transport and generation of bound states},}\ }\href
  {\doibase 10.1103/PhysRevB.96.104309} {\bibfield  {journal} {\bibinfo
  {journal} {Phys. Rev. B}\ }\textbf {\bibinfo {volume} {96}},\ \bibinfo
  {pages} {104309} (\bibinfo {year} {2017})}\BibitemShut {NoStop}%
\bibitem [{\citenamefont {H\"ubner}\ \emph {et~al.}(2022)\citenamefont
  {H\"ubner}, \citenamefont {Dauer}, \citenamefont {Eggert}, \citenamefont
  {Kollath},\ and\ \citenamefont {Sheikhan}}]{HubnerLocalDrive}%
  \BibitemOpen
  \bibfield  {author} {\bibinfo {author} {\bibfnamefont {Friedrich}\
  \bibnamefont {H\"ubner}}, \bibinfo {author} {\bibfnamefont {Christoph}\
  \bibnamefont {Dauer}}, \bibinfo {author} {\bibfnamefont {Sebastian}\
  \bibnamefont {Eggert}}, \bibinfo {author} {\bibfnamefont {Corinna}\
  \bibnamefont {Kollath}}, \ and\ \bibinfo {author} {\bibfnamefont {Ameneh}\
  \bibnamefont {Sheikhan}},\ }\bibfield  {title} {\enquote {\bibinfo {title}
  {Floquet-engineered pair and single-particle filters in the fermi-hubbard
  model},}\ }\href {\doibase 10.1103/PhysRevA.106.043303} {\bibfield  {journal}
  {\bibinfo  {journal} {Phys. Rev. A}\ }\textbf {\bibinfo {volume} {106}},\
  \bibinfo {pages} {043303} (\bibinfo {year} {2022})}\BibitemShut {NoStop}%
\bibitem [{\citenamefont {Bethe}(1931)}]{Bethe_1931}%
  \BibitemOpen
  \bibfield  {author} {\bibinfo {author} {\bibfnamefont {H.}~\bibnamefont
  {Bethe}},\ }\bibfield  {title} {\enquote {\bibinfo {title} {Zur theorie der
  metalle},}\ }\href@noop {} {\bibfield  {journal} {\bibinfo  {journal}
  {Zeitschrift f{\"u}r Physik}\ }\textbf {\bibinfo {volume} {71}},\ \bibinfo
  {pages} {205--226} (\bibinfo {year} {1931})}\BibitemShut {NoStop}%
\bibitem [{\citenamefont {Mukherjee}\ \emph {et~al.}(2024)\citenamefont
  {Mukherjee}, \citenamefont {Melendrez}, \citenamefont {Szyniszewski},
  \citenamefont {Changlani},\ and\ \citenamefont
  {Pal}}]{Mukherjee_Melendrez_2024}%
  \BibitemOpen
  \bibfield  {author} {\bibinfo {author} {\bibfnamefont {Bhaskar}\ \bibnamefont
  {Mukherjee}}, \bibinfo {author} {\bibfnamefont {Ronald}\ \bibnamefont
  {Melendrez}}, \bibinfo {author} {\bibfnamefont {Marcin}\ \bibnamefont
  {Szyniszewski}}, \bibinfo {author} {\bibfnamefont {Hitesh~J.}\ \bibnamefont
  {Changlani}}, \ and\ \bibinfo {author} {\bibfnamefont {Arijeet}\ \bibnamefont
  {Pal}},\ }\bibfield  {title} {\enquote {\bibinfo {title} {Emergent strong
  zero mode through local floquet engineering},}\ }\href {\doibase
  10.1103/PhysRevB.109.064303} {\bibfield  {journal} {\bibinfo  {journal}
  {Phys. Rev. B}\ }\textbf {\bibinfo {volume} {109}},\ \bibinfo {pages}
  {064303} (\bibinfo {year} {2024})}\BibitemShut {NoStop}%
\bibitem [{\citenamefont {Haldane}(1983)}]{HaldaneSpin1}%
  \BibitemOpen
  \bibfield  {author} {\bibinfo {author} {\bibfnamefont {F.~D.~M.}\
  \bibnamefont {Haldane}},\ }\bibfield  {title} {\enquote {\bibinfo {title}
  {Nonlinear field theory of large-spin heisenberg antiferromagnets:
  Semiclassically quantized solitons of the one-dimensional easy-axis n\'eel
  state},}\ }\href {\doibase 10.1103/PhysRevLett.50.1153} {\bibfield  {journal}
  {\bibinfo  {journal} {Phys. Rev. Lett.}\ }\textbf {\bibinfo {volume} {50}},\
  \bibinfo {pages} {1153--1156} (\bibinfo {year} {1983})}\BibitemShut {NoStop}%
\bibitem [{\citenamefont {Hida}(1992)}]{Hida1992AlternatingChain}%
  \BibitemOpen
  \bibfield  {author} {\bibinfo {author} {\bibfnamefont {Kazuo}\ \bibnamefont
  {Hida}},\ }\bibfield  {title} {\enquote {\bibinfo {title} {Crossover between
  the haldane-gap phase and the dimer phase in the spin-1/2 alternating
  heisenberg chain},}\ }\href {\doibase 10.1103/PhysRevB.45.2207} {\bibfield
  {journal} {\bibinfo  {journal} {Phys. Rev. B}\ }\textbf {\bibinfo {volume}
  {45}},\ \bibinfo {pages} {2207--2212} (\bibinfo {year} {1992})}\BibitemShut
  {NoStop}%
\bibitem [{\citenamefont {Kohmoto}\ and\ \citenamefont
  {Tasaki}(1992)}]{KohmotoAlternating}%
  \BibitemOpen
  \bibfield  {author} {\bibinfo {author} {\bibfnamefont {Mahito}\ \bibnamefont
  {Kohmoto}}\ and\ \bibinfo {author} {\bibfnamefont {Hal}\ \bibnamefont
  {Tasaki}},\ }\bibfield  {title} {\enquote {\bibinfo {title} {Hidden
  ${\mathit{z}}_{2}$\ifmmode\times\else\texttimes\fi{}${\mathit{z}}_{2}$
  symmetry breaking and the haldane phase in the s=1/2 quantum spin chain with
  bond alternation},}\ }\href {\doibase 10.1103/PhysRevB.46.3486} {\bibfield
  {journal} {\bibinfo  {journal} {Phys. Rev. B}\ }\textbf {\bibinfo {volume}
  {46}},\ \bibinfo {pages} {3486--3495} (\bibinfo {year} {1992})}\BibitemShut
  {NoStop}%
\bibitem [{\citenamefont {Jacquod}\ \emph {et~al.}(2001)\citenamefont
  {Jacquod}, \citenamefont {Silvestrov},\ and\ \citenamefont
  {Beenakker}}]{BennakerKicked}%
  \BibitemOpen
  \bibfield  {author} {\bibinfo {author} {\bibfnamefont {Ph.}\ \bibnamefont
  {Jacquod}}, \bibinfo {author} {\bibfnamefont {P.G.}\ \bibnamefont
  {Silvestrov}}, \ and\ \bibinfo {author} {\bibfnamefont {C.W.J.}\ \bibnamefont
  {Beenakker}},\ }\bibfield  {title} {\enquote {\bibinfo {title} {Golden rule
  decay versus lyapunov decay of the quantum loschmidt echo},}\ }\href
  {\doibase 10.1103/PhysRevE.64.055203} {\bibfield  {journal} {\bibinfo
  {journal} {Phys. Rev. E}\ }\textbf {\bibinfo {volume} {64}},\ \bibinfo
  {pages} {055203} (\bibinfo {year} {2001})}\BibitemShut {NoStop}%
\bibitem [{\citenamefont {Haake}\ \emph {et~al.}(1987)\citenamefont {Haake},
  \citenamefont {Ku{\'s}},\ and\ \citenamefont {Scharf}}]{haake1987classical}%
  \BibitemOpen
  \bibfield  {author} {\bibinfo {author} {\bibfnamefont {Fritz}\ \bibnamefont
  {Haake}}, \bibinfo {author} {\bibfnamefont {M}~\bibnamefont {Ku{\'s}}}, \
  and\ \bibinfo {author} {\bibfnamefont {Rainer}\ \bibnamefont {Scharf}},\
  }\bibfield  {title} {\enquote {\bibinfo {title} {Classical and quantum chaos
  for a kicked top},}\ }\href@noop {} {\bibfield  {journal} {\bibinfo
  {journal} {Zeitschrift f{\"u}r Physik B Condensed Matter}\ }\textbf {\bibinfo
  {volume} {65}},\ \bibinfo {pages} {381--395} (\bibinfo {year}
  {1987})}\BibitemShut {NoStop}%
\bibitem [{\citenamefont {Sinha}\ \emph {et~al.}(2021)\citenamefont {Sinha},
  \citenamefont {Ray},\ and\ \citenamefont {Sinha}}]{Sinha_2021KickedDicke}%
  \BibitemOpen
  \bibfield  {author} {\bibinfo {author} {\bibfnamefont {Sudip}\ \bibnamefont
  {Sinha}}, \bibinfo {author} {\bibfnamefont {Sayak}\ \bibnamefont {Ray}}, \
  and\ \bibinfo {author} {\bibfnamefont {Subhasis}\ \bibnamefont {Sinha}},\
  }\bibfield  {title} {\enquote {\bibinfo {title} {Fingerprint of chaos and
  quantum scars in kicked dicke model: an out-of-time-order correlator
  study},}\ }\href {\doibase 10.1088/1361-648X/abe26b} {\bibfield  {journal}
  {\bibinfo  {journal} {Journal of Physics: Condensed Matter}\ }\textbf
  {\bibinfo {volume} {33}},\ \bibinfo {pages} {174005} (\bibinfo {year}
  {2021})}\BibitemShut {NoStop}%
\bibitem [{\citenamefont {Vidal}(2003)}]{Vidal_TEBD}%
  \BibitemOpen
  \bibfield  {author} {\bibinfo {author} {\bibfnamefont {Guifré}\ \bibnamefont
  {Vidal}},\ }\bibfield  {title} {\enquote {\bibinfo {title} {Efficient
  classical simulation of slightly entangled quantum computations},}\ }\href
  {\doibase 10.1103/physrevlett.91.147902} {\bibfield  {journal} {\bibinfo
  {journal} {Physical Review Letters}\ }\textbf {\bibinfo {volume} {91}}
  (\bibinfo {year} {2003}),\ 10.1103/physrevlett.91.147902}\BibitemShut
  {NoStop}%
\bibitem [{Note1()}]{Note1}%
  \BibitemOpen
  \bibinfo {note} {We also note that similar observations, albeit for a
  different model and observables, have been reported recently in Ref.~\cite
  {Duan_scar}.}\BibitemShut {Stop}%
\bibitem [{\citenamefont {Abanin}\ \emph {et~al.}(2017)\citenamefont {Abanin},
  \citenamefont {De~Roeck}, \citenamefont {Ho},\ and\ \citenamefont
  {Huveneers}}]{Abanin2017}%
  \BibitemOpen
  \bibfield  {author} {\bibinfo {author} {\bibfnamefont {Dmitry}\ \bibnamefont
  {Abanin}}, \bibinfo {author} {\bibfnamefont {Wojciech}\ \bibnamefont
  {De~Roeck}}, \bibinfo {author} {\bibfnamefont {Wen~Wei}\ \bibnamefont {Ho}},
  \ and\ \bibinfo {author} {\bibfnamefont {Fran{\c{c}}ois}\ \bibnamefont
  {Huveneers}},\ }\bibfield  {title} {\enquote {\bibinfo {title} {A rigorous
  theory of many-body prethermalization for periodically driven and closed
  quantum systems},}\ }\href {\doibase 10.1007/s00220-017-2930-x} {\bibfield
  {journal} {\bibinfo  {journal} {Communications in Mathematical Physics}\
  }\textbf {\bibinfo {volume} {354}},\ \bibinfo {pages} {809--827} (\bibinfo
  {year} {2017})}\BibitemShut {NoStop}%
\bibitem [{\citenamefont {Wen}\ \emph {et~al.}(2022)\citenamefont {Wen},
  \citenamefont {Fan},\ and\ \citenamefont {Vishwanath}}]{Wen_2022}%
  \BibitemOpen
  \bibfield  {author} {\bibinfo {author} {\bibfnamefont {Xueda}\ \bibnamefont
  {Wen}}, \bibinfo {author} {\bibfnamefont {Ruihua}\ \bibnamefont {Fan}}, \
  and\ \bibinfo {author} {\bibfnamefont {Ashvin}\ \bibnamefont {Vishwanath}},\
  }\href {\doibase 10.48550/ARXIV.2211.00040} {\enquote {\bibinfo {title}
  {Floquet's refrigerator: Conformal cooling in driven quantum critical
  systems},}\ } (\bibinfo {year} {2022})\BibitemShut {NoStop}%
\bibitem [{\citenamefont {Su}\ \emph {et~al.}()\citenamefont {Su},
  \citenamefont {Sun}, \citenamefont {Hudomal}, \citenamefont {Desaules},
  \citenamefont {Zhou}, \citenamefont {Yang}, \citenamefont {Halimeh},
  \citenamefont {Yuan}, \citenamefont {Papić},\ and\ \citenamefont
  {Pan}}]{Jad}%
  \BibitemOpen
  \bibfield  {author} {\bibinfo {author} {\bibfnamefont {Guo-Xian}\
  \bibnamefont {Su}}, \bibinfo {author} {\bibfnamefont {Hui}\ \bibnamefont
  {Sun}}, \bibinfo {author} {\bibfnamefont {Ana}\ \bibnamefont {Hudomal}},
  \bibinfo {author} {\bibfnamefont {Jean-Yves}\ \bibnamefont {Desaules}},
  \bibinfo {author} {\bibfnamefont {Zhao-Yu}\ \bibnamefont {Zhou}}, \bibinfo
  {author} {\bibfnamefont {Bing}\ \bibnamefont {Yang}}, \bibinfo {author}
  {\bibfnamefont {Jad~C.}\ \bibnamefont {Halimeh}}, \bibinfo {author}
  {\bibfnamefont {Zhen-Sheng}\ \bibnamefont {Yuan}}, \bibinfo {author}
  {\bibfnamefont {Zlatko}\ \bibnamefont {Papić}}, \ and\ \bibinfo {author}
  {\bibfnamefont {Jian-Wei}\ \bibnamefont {Pan}},\ }\href@noop {} {\enquote
  {\bibinfo {title} {Observation of unconventional many-body scarring in a
  quantum simulator},}\ }\Eprint {http://arxiv.org/abs/2201.00821}
  {arXiv:2201.00821} \BibitemShut {NoStop}%
\bibitem [{\citenamefont {Tan}\ \emph {et~al.}(2021)\citenamefont {Tan},
  \citenamefont {Becker}, \citenamefont {Liu}, \citenamefont {Pagano},
  \citenamefont {Collins}, \citenamefont {De}, \citenamefont {Feng},
  \citenamefont {Kaplan}, \citenamefont {Kyprianidis}, \citenamefont
  {Lundgren}, \citenamefont {Morong}, \citenamefont {Whitsitt}, \citenamefont
  {Gorshkov},\ and\ \citenamefont {Monroe}}]{Tan_TFIM_2021}%
  \BibitemOpen
  \bibfield  {author} {\bibinfo {author} {\bibfnamefont {W.~L.}\ \bibnamefont
  {Tan}}, \bibinfo {author} {\bibfnamefont {P.}~\bibnamefont {Becker}},
  \bibinfo {author} {\bibfnamefont {F.}~\bibnamefont {Liu}}, \bibinfo {author}
  {\bibfnamefont {G.}~\bibnamefont {Pagano}}, \bibinfo {author} {\bibfnamefont
  {K.~S.}\ \bibnamefont {Collins}}, \bibinfo {author} {\bibfnamefont
  {A.}~\bibnamefont {De}}, \bibinfo {author} {\bibfnamefont {L.}~\bibnamefont
  {Feng}}, \bibinfo {author} {\bibfnamefont {H.~B.}\ \bibnamefont {Kaplan}},
  \bibinfo {author} {\bibfnamefont {A.}~\bibnamefont {Kyprianidis}}, \bibinfo
  {author} {\bibfnamefont {R.}~\bibnamefont {Lundgren}}, \bibinfo {author}
  {\bibfnamefont {W.}~\bibnamefont {Morong}}, \bibinfo {author} {\bibfnamefont
  {S.}~\bibnamefont {Whitsitt}}, \bibinfo {author} {\bibfnamefont {A.~V.}\
  \bibnamefont {Gorshkov}}, \ and\ \bibinfo {author} {\bibfnamefont
  {C.}~\bibnamefont {Monroe}},\ }\bibfield  {title} {\enquote {\bibinfo {title}
  {Domain-wall confinement and dynamics in a quantum simulator},}\ }\href
  {\doibase 10.1038/s41567-021-01194-3} {\bibfield  {journal} {\bibinfo
  {journal} {Nature Physics}\ }\textbf {\bibinfo {volume} {17}},\ \bibinfo
  {pages} {742--747} (\bibinfo {year} {2021})}\BibitemShut {NoStop}%
\bibitem [{\citenamefont {Jepsen}\ \emph {et~al.}(2022)\citenamefont {Jepsen},
  \citenamefont {Lee}, \citenamefont {Lin}, \citenamefont {Dimitrova},
  \citenamefont {Margalit}, \citenamefont {Ho},\ and\ \citenamefont
  {Ketterle}}]{Jepsen2022}%
  \BibitemOpen
  \bibfield  {author} {\bibinfo {author} {\bibfnamefont {Paul~Niklas}\
  \bibnamefont {Jepsen}}, \bibinfo {author} {\bibfnamefont {Yoo
  Kyung~`Eunice'}\ \bibnamefont {Lee}}, \bibinfo {author} {\bibfnamefont
  {Hanzhen}\ \bibnamefont {Lin}}, \bibinfo {author} {\bibfnamefont {Ivana}\
  \bibnamefont {Dimitrova}}, \bibinfo {author} {\bibfnamefont {Yair}\
  \bibnamefont {Margalit}}, \bibinfo {author} {\bibfnamefont {Wen~Wei}\
  \bibnamefont {Ho}}, \ and\ \bibinfo {author} {\bibfnamefont {Wolfgang}\
  \bibnamefont {Ketterle}},\ }\bibfield  {title} {\enquote {\bibinfo {title}
  {Long-lived phantom helix states in heisenberg quantum magnets},}\ }\href
  {\doibase 10.1038/s41567-022-01651-7} {\bibfield  {journal} {\bibinfo
  {journal} {Nature Physics}\ }\textbf {\bibinfo {volume} {18}},\ \bibinfo
  {pages} {899--904} (\bibinfo {year} {2022})}\BibitemShut {NoStop}%
\bibitem [{\citenamefont {Fishman}\ \emph {et~al.}(2020)\citenamefont
  {Fishman}, \citenamefont {White},\ and\ \citenamefont
  {Stoudenmire}}]{ITensor}%
  \BibitemOpen
  \bibfield  {author} {\bibinfo {author} {\bibfnamefont {Matthew}\ \bibnamefont
  {Fishman}}, \bibinfo {author} {\bibfnamefont {Steven~R.}\ \bibnamefont
  {White}}, \ and\ \bibinfo {author} {\bibfnamefont {E.~Miles}\ \bibnamefont
  {Stoudenmire}},\ }\href@noop {} {\enquote {\bibinfo {title} {The
  \mbox{ITensor} software library for tensor network calculations},}\ }
  (\bibinfo {year} {2020}),\ \Eprint {http://arxiv.org/abs/2007.14822}
  {arXiv:2007.14822} \BibitemShut {NoStop}%
\bibitem [{ase(2021)}]{asen}%
  \BibitemOpen
  \bibfield  {title} {\enquote {\bibinfo {title} {Analytic approaches to
  periodically driven closed quantum systems: methods and applications},}\
  }\href {\doibase 10.1088/1361-648X/ac1b61} {\bibfield  {journal} {\bibinfo
  {journal} {Journal of Physics: Condensed Matter}\ }\textbf {\bibinfo {volume}
  {33}} (\bibinfo {year} {2021}),\ 10.1088/1361-648X/ac1b61}\BibitemShut
  {NoStop}%
\bibitem [{\citenamefont {Vajna}\ \emph {et~al.}(2018)\citenamefont {Vajna},
  \citenamefont {Klobas}, \citenamefont {Prosen},\ and\ \citenamefont
  {Polkovnikov}}]{replica}%
  \BibitemOpen
  \bibfield  {author} {\bibinfo {author} {\bibfnamefont {Szabolcs}\
  \bibnamefont {Vajna}}, \bibinfo {author} {\bibfnamefont {Katja}\ \bibnamefont
  {Klobas}}, \bibinfo {author} {\bibfnamefont {Toma\ifmmode
  \check{z}\else~\v{z}\fi{}}\ \bibnamefont {Prosen}}, \ and\ \bibinfo {author}
  {\bibfnamefont {Anatoli}\ \bibnamefont {Polkovnikov}},\ }\bibfield  {title}
  {\enquote {\bibinfo {title} {Replica resummation of the
  baker-campbell-hausdorff series},}\ }\href {\doibase
  10.1103/PhysRevLett.120.200607} {\bibfield  {journal} {\bibinfo  {journal}
  {Phys. Rev. Lett.}\ }\textbf {\bibinfo {volume} {120}},\ \bibinfo {pages}
  {200607} (\bibinfo {year} {2018})}\BibitemShut {NoStop}%
\bibitem [{\citenamefont {Bukov}\ \emph {et~al.}(2015)\citenamefont {Bukov},
  \citenamefont {D'Alessio},\ and\ \citenamefont {Polkovnikov}}]{anatoli1}%
  \BibitemOpen
  \bibfield  {author} {\bibinfo {author} {\bibfnamefont {Marin}\ \bibnamefont
  {Bukov}}, \bibinfo {author} {\bibfnamefont {Luca}\ \bibnamefont {D'Alessio}},
  \ and\ \bibinfo {author} {\bibfnamefont {Anatoli}\ \bibnamefont
  {Polkovnikov}},\ }\bibfield  {title} {\enquote {\bibinfo {title} {Universal
  high-frequency behavior of periodically driven systems: from dynamical
  stabilization to floquet engineering},}\ }\href {\doibase
  10.1080/00018732.2015.1055918} {\bibfield  {journal} {\bibinfo  {journal}
  {Advances in Physics}\ }\textbf {\bibinfo {volume} {64}},\ \bibinfo {pages}
  {139--226} (\bibinfo {year} {2015})},\ \Eprint
  {http://arxiv.org/abs/https://doi.org/10.1080/00018732.2015.1055918}
  {https://doi.org/10.1080/00018732.2015.1055918} \BibitemShut {NoStop}%
\bibitem [{\citenamefont {D’Alessio}\ and\ \citenamefont
  {Polkovnikov}(2013)}]{anatoli2}%
  \BibitemOpen
  \bibfield  {author} {\bibinfo {author} {\bibfnamefont {Luca}\ \bibnamefont
  {D’Alessio}}\ and\ \bibinfo {author} {\bibfnamefont {Anatoli}\ \bibnamefont
  {Polkovnikov}},\ }\bibfield  {title} {\enquote {\bibinfo {title} {Many-body
  energy localization transition in periodically driven systems},}\ }\href
  {\doibase https://doi.org/10.1016/j.aop.2013.02.011} {\bibfield  {journal}
  {\bibinfo  {journal} {Annals of Physics}\ }\textbf {\bibinfo {volume}
  {333}},\ \bibinfo {pages} {19--33} (\bibinfo {year} {2013})}\BibitemShut
  {NoStop}%
\bibitem [{\citenamefont {Yuan}\ \emph {et~al.}(2022)\citenamefont {Yuan},
  \citenamefont {Zhang}, \citenamefont {Wang}, \citenamefont {Duan},\ and\
  \citenamefont {Deng}}]{Duan_scar}%
  \BibitemOpen
  \bibfield  {author} {\bibinfo {author} {\bibfnamefont {Dong}\ \bibnamefont
  {Yuan}}, \bibinfo {author} {\bibfnamefont {Shun-Yao}\ \bibnamefont {Zhang}},
  \bibinfo {author} {\bibfnamefont {Yu}~\bibnamefont {Wang}}, \bibinfo {author}
  {\bibfnamefont {L.-M.}\ \bibnamefont {Duan}}, \ and\ \bibinfo {author}
  {\bibfnamefont {Dong-Ling}\ \bibnamefont {Deng}},\ }\bibfield  {title}
  {\enquote {\bibinfo {title} {Quantum information scrambling in quantum
  many-body scarred systems},}\ }\href {\doibase
  10.1103/PhysRevResearch.4.023095} {\bibfield  {journal} {\bibinfo  {journal}
  {Phys. Rev. Res.}\ }\textbf {\bibinfo {volume} {4}},\ \bibinfo {pages}
  {023095} (\bibinfo {year} {2022})}\BibitemShut {NoStop}%
\end{thebibliography}%

\end{document}